\documentclass[a4paper,12pt]{article}
\usepackage{amssymb,amsmath,mathbbol,mathrsfs}
\usepackage[usenames,dvipsnames]{color}
\usepackage{hyperref}
\usepackage{xcolor}
\usepackage{stmaryrd}
\usepackage{authblk}
\usepackage{framed}
\usepackage{empheq} 
\usepackage{slashed}
\usepackage{hyphenat}
\usepackage{cite}
\usepackage{apacite}
\usepackage{natbib}
\usepackage{dsfont}
\usepackage{chngcntr}
\usepackage{enumerate}  

\usepackage{marginnote}    

\usepackage[left=.8in,right=.8in,top=.8in,bottom=.8in]{geometry}                  % For good copy.

\usepackage{sectsty} 
\allsectionsfont{\sffamily\mdseries\upshape} % (See the fntguide.pdf for font help)
\usepackage{tocloft}

\makeatletter
\renewenvironment{abstract}{%
    \if@twocolumn
      \section*{\abstractname}%
    \else %% <- here I've removed \small
      \begin{center}%
        {\bfseries\sffamily\abstractname\vspace{\z@}}%  %% <- here I've added \Large
      \end{center}%
      \quotation
    \fi}
    {\if@twocolumn\else\endquotation\fi}
\makeatother

\numberwithin{equation}{section}
\hypersetup{
	colorlinks=true,         
	linkcolor=MidnightBlue,          
	citecolor=BrickRed,        
	urlcolor=MidnightBlue            
}

\newcommand{\be}{\begin{equation}}
\newcommand{\ee}{\end{equation}}

\renewcommand{\d}{{\mathrm{d}}}
\newcommand{\D}{{\mathrm{D}}}

\newcommand{\Ad}{{\mathrm{Ad}}}
\newcommand{\pp}{{\partial}}

\renewcommand{\hat}{\widehat}

\newcommand{\RR}{\mathds{R}} %Isham-style

\newtheorem{defi}{Definition}

\newcommand{\cint}{{\int\kern-.87em{<}}}
\newcommand{\sint}{{\int\kern-.75em{\sim}}}
\newcommand{\fint}{{\int\kern-1.00em{\int}}}

\newcommand{\bb}{\mathbb}

\renewcommand{\#}{\sharp}
\let\oldmarginpar\marginpar
\renewcommand\marginpar[1]{\oldmarginpar{\color{red}\raggedright\footnotesize #1}}

\begin{document}

\title{The Aharonov-Bohm effect: reality and folklore}
\author{Henrique Gomes\\ 
Oriel College, University of Oxford}

\maketitle

\begin{abstract} 
The Aharonov-Bohm (A-B) effect has been a major focus of the foundations of physics. And yet, much confusion persists. In particular, the effect purportedly leads to a dilemma: on one horn, we have a non-local action of a gauge-invariant quantity on charged particles; on the other, we get a local action on these particles, but of a non-gauge invariant quantity.  This is the folklore, but the folklore is filled with misconceptions. Here, by deploying a recently defended formulation of gauge theory that dispenses with principal bundles, gauge potentials, and explicit gauge symmetries, I argue, with previous authors, that the A-B effect can be understood gauge-independently. But here my argument will go further: I will show that the A-B effect, when expressed in terms of the covariant derivative of a vector bundle, is \emph{entirely} analogous to the holonomy of \emph{spacetime} vectors, and can be understood completely locally.  The only surprising idea illustrated by the A-B effect is that, in some circumstances, there is more to the covariant derivative than can be accounted for by the curvature and underlying topology of a vector bundle.
\end{abstract}

\tableofcontents
\section{Introduction}
It is often said that the Aharonov-Bohm effect of 
electromagnetism---henceforth, the A-B effect---evinces a form of non-locality, something
that is usually thought of as confined to nonclassical physics (cf e.g. \citep{Myrvold2010, sep-physics-holism, Belot1998, Healey_book}).  Thus the effect is often portrayed both as having a quantum nature and as relying on the non-trivial topology of space, which is an obvious non-local fact. For instance, \cite{healey1997ab} says it is a kind of non-locality \lq\lq{}much more closely analogous to the kind of nonlocality manifested by a violation of Bell inequalities than has been previously acknowledged\rq\rq{}. 

But I disagree with all three elements of this portrayal: (i) that the A-B effect has a quantum nature; (ii) that it relies on the non-trivial topology of space; and (iii) that it evinces an important form of non-locality. While none of these claims are wholly unchallenged in the literature, they remain widespread—especially in philosophical discussions—as a kind of conceptual shorthand. My aim is to dislodge this received picture---the \lq{}folklore\rq{} if you will---and to offer in its place a more coherent and conceptually unified account of the A–B effect.% I am  not the first to disagree with any of (i-iii); nonetheless they remain entrenched as folklore: \lq\lq{}traditional stories of a group of people\rq\rq{}---in this case, the group is of physicists and philosophers. I hope to finally dislodge this folklore with a rebuttal of (i-iii) grounded on a tight analogy between the A-B effect and the parallel transport of spacetime vectors.  

This alternative account is structured around a single unifying claim: the A–B effect is a special case of parallel transport around a loop, and thus entirely analogous to the holonomy of spacetime vectors in general relativity. That is, the effect can be understood geometrically as a feature of the covariant derivative associated with a connection on a vector bundle—without reference to gauge potentials, quantum entanglement, or exotic topologies.\footnote{ Within the literature, as far as I know, \cite{Leeds1999}  is the only author who attempted to use a similar view of gauge theory to dissolve the puzzle of the A-B effect. Unfortunately, his attempt focused on the metaphysical implications of the view and lacked mathematical precision. This lack of precision led to an equivocation between different concepts (ibid, p. 614) which was picked up by \cite[Sec. 4.2.2]{Healey_book}, shown to lead to inconsistencies, and finally used to dismiss the approach. In the final Section \ref{sec:conclusions}, I will briefly revisit this episode in the debate surrounding the A-B effect. }

This understanding becomes especially sharp when we reformulate gauge theory using vector bundles and covariant derivatives, dispensing with the apparatus of principal bundles and gauge potentials. In this alternative formulation, developed in \citep{Gomes_internal}, all the fundamental interactions acquire the same geometrical nature: they are all based on covariant derivatives describing the geometry of vector bundles.  The formulation will be introduced in Section \ref{sec:vector}, but its conceptual significance informs the argument throughout.

   I will argue that the type of non-locality at play in the A-B effect is relatively benign and pervasive in physics. It arises from the comparison of quantities that depend on their spacetime history. Although broadly similar arguments to this effect have been made before,  my argument will be grounded on  a tight analogy between the A-B effect and the holonomy of a spacetime vector.\footnote{This mathematical object will be introduced formally in Section \ref{sec:vector} (e.g. Equation \eqref{eq:PT} for the holonomy in an arbitrary vector bundle). Until then, I will mention terms like \lq{}holonomy, covariant derivative, and curvature\rq{} in the spacetime context, which I assume to be understood. On a Lorentian spacetime $(M, g_{ab})$ the Levi-Civita covariant derivative $\nabla$ is an operator $\nabla:TM\otimes \Gamma(TM)\rightarrow \Gamma(TM)$ (which is $\RR$-linear in its first entry and $C^\infty(M)$ linear on its second), which defines the infinitesimal parallel transport of vector fields via its kernel (see Equation \ref{ftnt:TE}). Holonomy is an operator that takes parallel transport along closed curves $\gamma: S^1\rightarrow M$ and, for $x\in \gamma$, gives an endomorphism $\mathsf{End}(T_xM)$, i.e. a linear transformation at $T_xM$ that preserves whatever structure $\nabla$ does. The curvature associated to $\nabla$ is a $C^\infty(M)$-multi-linear operator $\Omega : TM \otimes TM \otimes TM \to TM$ defined as $\Omega=  \nabla_X \nabla_Y - \nabla_Y \nabla_X - \nabla_{[X,Y]}$ that results from taking an infinitesimal holonomy (just like the covariant derivative results from taking the infinitesimal parallel transport).  these mathematical objects exist for any vector bundle, as we will see.   } 

Beyond correcting what I take to be common misconceptions, I will argue that the A-B effect and its kin expose physically salient features of the covariant derivative that curvature and topology alone fail to capture. The effect illustrates something  important and often overlooked: the fundamental significance of the affine structure encoded by the covariant derivative, beyond what is encoded in the curvature and topology.

 Here is how I plan to proceed. In Section \ref{sec:curv_conn},  I introduce the standard setup of the A–B effect and recast it in geometric terms, showing that the effect probes a purely classical feature of electromagnetism. I also exhibit a non-Abelian analogue of the effect in a topologically trivial setting, undermining the supposed necessity of non-trivial topology. In Section \ref{sec:grav_analog}  I develop two gravitational analogies: the first highlights a disanalogy between the A–B effect and proper time accumulation in the context of the twin paradox (cf. \citep{Jacobs_PFB}), while the second shows that the A–B effect is structurally identical to the holonomy of parallel-transported vectors in GR.  Section \ref{sec:vector} introduces the vector bundle point of view of gauge theory, and shows how it captures the explanatory core of the A–B effect in a way that is fully local, classical, and geometrically transparent. In Section \ref{sec:critic}, I respond to possible criticisms of the claims being made in favor of the vector bundle point of view. In Section \ref{sec:conclusions} I conclude by summarising.
%Historically, it was Faraday who discovered, in 1831, that a magnetic field varying in time would induce a non-zero curl in the electric field (\ref{conservcurr}). In 1861, Maxwell realized that the law of conservation of (electric) charges could be automatically introduced into the Biot-Savart Law: $$\nabla\times\vec B=\vec \jmath$$ By simply adding to it a term $-\frac{\partial \vec E}{\partial t}=\vec \jmath$ and taking the divergence, he arrived at the continuity equation \be\label{continuidade}\frac{d \rho}{d t}=\nabla\cdot\vec\jmath\ee

\section{The A-B effect}\label{sec:curv_conn}\label{sec:A-B}
%Does  the physical content of the gauge potential in the Abelian theory outstrip that of the Maxwell Faraday tensor? As is immediate to observe from \eqref{eq:cov_curv}, the curvature is gauge-invariant in the Abelian case. This often leads to questions about whether physical theories couldn't be entirely described without the use of the gauge \emph{variant} potentials. But surprisingly, Abelian gauge theory has more than  curvature as its fundamental degrees of freedom. The A-B effect  describes physical, or gauge-invariant,  features of the theory that cannot be articulated using only the curvature. These features appear even in vacuum, though there they require spacetime to have (effectively) a non-trivial topology. %I will discuss this standard presentation of the A-B effect in Section \ref{sec:A-B}. 

In Section \ref{sec:setup} I will first present the standard setup of the A-B effect and the dillemma about locality and gauge invariance that it poses. Then in Section \ref{sec:AB_EM} I will give a geometric description of the effect and argue that it can be understood as saying something purely about the classical theory of electromagnetism. In Section \ref{sec:non_Ab} I will give a closely analogous effect in the non-Abelian case, which, unlike the Abelian case, works for a trivial topological background. 

\subsection{Setup}\label{sec:setup}
Classically, the magnetic field \( \mathbf{B} \) is typically associated with the gauge potential \( \mathbf{A} \) through the relation
\be
\mathbf{B} = \nabla \times \mathbf{A}.
\ee
From this relation, it is clear that a transformation $\mathbf{A}\rightarrow \mathbf{A}+\nabla \lambda$, for $\lambda$ any real-valued function, doesn\rq{}t change the magnetic field. Indeed, this transformation of the potential is called a \emph{gauge transformation}, and it is taken to leave all physical quantities invariant.\footnote{Because in this case the transformations are commutative---we obtain the same result transforming $\mathbf{A}$ using $\lambda_1$ and then transforming the result using $\lambda_2$ as if we do it in the opposite order---the gauge theory in question is \emph{Abelian}.} That the transformation is redundant underscores the idea the gauge potential is merely a computational tool, with all physical content encoded entirely in the magnetic field.

Historically,  Aharonov and Bohm put this interpretation of $\mathbf{A}$ under pressure by proposing an electron interference experiment, in which a beam is split into two branches which go around a solenoid and are brought back together to form an interference pattern.\footnote{\cite{aharonovbohm1959}'s work was  conducted independently of the work by \citet{ehrenberg1949refractive} who proposed the same experiment with a different framing in a work that did not receive much attention at the time. According to \citet{hiley2013ABeffect}, the effect was discovered ``at least three times before Aharonov and Bohm's paper''; with the first being a talk by Walter Franz, which described a similar experiment in 1939.} A solenoid is a conducting
wire coiled around a cylinder whose length is long compared to the wavelength
of the particle under consideration. When current runs through the
solenoid, a magnetic field is created inside the device, but the external
magnetic field is unaffected. In other words, this solenoid is
perfectly shielded, so that the magnetic field vanishes outside it, and our model makes the simplifying assumption that no electron can penetrate inside and detect the magnetic field directly.

\begin{figure}[h!]
\center
\includegraphics[width=0.4\textwidth]{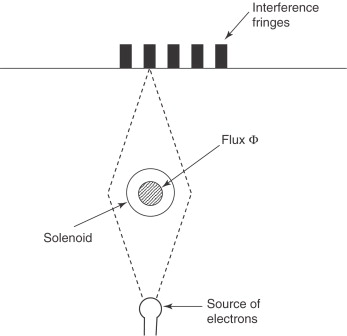}
\caption{The outlines of the experimental setup for detection of the A-B effect.}\label{fig1}
\end{figure}

The experiment involves two different configurations: solenoid on or off. In both, the field-strength (i.e. the magnetic field)  along the paths accessible to the charged particles is zero (see figure \ref{fig1}). The surprising---and confirmed!---prediction of electromagnetism, as fleshed out by Aharonov and Bohm, is that the two configurations produce two different interference patterns.  As
the magnetic flux in the solenoid changes, the interference fringes shift. 

The key insight of the A-B effect is that charged particles, such as electrons, can be influenced by the vector potential \( \mathbf{A} \) even in regions where the magnetic field strength vanishes,  \( \mathbf{B} = 0 \). %Specifically, electrons may experience a non-zero phase shift in their wavefunction \( \psi \) when encircling a region with a non-zero magnetic flux, despite the absence of any local magnetic field in the path of the particles.
The A-B effect suggests that the vector potential possesses a physical significance that transcends its role as a mathematical tool for describing the magnetic field. 

 So far, I agree. But there is a common conception that A-B effect also has profound implications for our understanding of locality vis-\`a-vis gauge invariance. With this, I will  disagree.

Those supposedly profound implications can be heuristically gleaned from my general outline of the experiment above. To recap:\\
\noindent (a) The observable phenomena change when the current in the solenoid changes; and\\
\noindent (b) The electrons that produce the phenomena are shielded from entering the region of non-zero magnetic fields; so\\
\noindent (c) If we rule out unmediated action-at-a-distance from the magnetic field,  whatever  physical difference  accounts for the change  must be due to differences in the gauge potential outside the solenoid. 

Thus, in order to explain the different patterns, one must either postulate a non-local action of the field-strength upon the particles, or endow the gauge potential with its own ontic significance, as producing a local physical effect on the electrons. 
Dismissing a non-local action singles out the second route. But that route  has problems.

 In order to describe the local effect of the gauge potential on the electron, we must choose one specific profile for the gauge potential, among infinitely many profiles that are gauge-related. But different profiles will tell  different stories about how the phases of the electrons were accrued along the trajectories; the accrual of phase has no gauge-invariant account. All roads seem to point to a gauge-invariant ontology that stubbornly resists a local account. So how can we resolve the puzzle posed by the A-B effect while preserving two cherished principles of modern physics: the locality and gauge-invariance of physical processes?

%We can simplify our treatment by  assuming throughout that there is no background charge nor field on a non-simply connected space $\Sigma=\RR^3-o$ where $o$ is the tube of the solenoid; and by taking the electrostatic situation,  considering only the spatial configuration of the fields. In this case, we identify the purely spatial component  of the Faraday tensor with the magnetic field and our electrons are confined to  travel within $\Sigma$.  

\subsection{The A-B effect probes a classical physical property}\label{sec:AB_EM}  
Here  I will argue that while experiments testing the A-B effect involve quantum mechanics, one can also witness the effect at a purely classical level.  (See \cite{Belot1998} for a more thorough philosophical analysis of the thesis defended in this section, and \citep{Stachel1982} for a technically precise defense of this thesis via a gravitational analogy.)

Supposing an electron is allowed to take any two  paths $\gamma_1$ and $\gamma_2^{-1}$ around the solenoid to the detector (it is convenient to parametrize the second path as going from the detector to the source).  We can infer from the shift in the
interference pattern that there are two contributions to
the relative phase of electron, corresponding to the paths that pass to the left and to the right of the solenoid. The final result is given by:\footnote{In units for which $e/\hbar c$=1. }
\be\label{eq:phase_A-B} e^{i\Delta}=\exp{\left(\left(i\int_{\gamma_1}\mathbf{A}\right)+\left(i\int_{\gamma^{-1}_2}\mathbf{A}\right)\right)}=\exp{\left(-i\oint_{\gamma_2\circ\gamma_1}\mathbf{A}\right)}, 
\ee

Now, I will assume the electrostatic situation, and that the bold-face $\mathbf{A}$ without indices denotes a spatial one-form. This one-form satisfies $\d\mathbf{A}=\mathbf{B}$, where $\d$ is the spatial exterior derivative.  A   gauge transformation  $\mathbf{A}\rightarrow \mathbf{A}+\d\lambda$ will not affect \eqref{eq:phase_A-B},  (for any $\lambda\in C^\infty(\Sigma)$), since $\gamma_2\circ\gamma_1\simeq S^1$, and so $\oint_{S^1}\d\lambda=0$,  by Stokes' theorem. Thus the phase difference $\Delta$ cares only about the gauge-equivalence class of $\mathbf A$.

%To relate this result with the discussions of Section \ref{ch:geom-I}:  the phase  cares only about the principal connection $\omega$, not about how we represent it on spacetime. And, since the magnetic field vanishes outside the solenoid in both situations, the connection $\omega$ is different in the two situations, although the curvature of that connection is the same, viz. zero.

The A-B effect shows  that the gauge potential cuts finer \emph{physical} distinctions than the field-strength tensor---also called the Faraday tensor, or the curvature tensor---can distinguish (within the regime discussed in this Section, this tensor is the magnetic field).  In other words, the A-B effect is about the physical significance of the gauge potential and the Faraday tensor, neither of which is being quantised in the above treatment. 
 In the case of electromagnetism, an Abelian ($U(1)$) gauge theory, we can find out precisely what is the physical information in the equivalence classes of the gauge potential that outstrips what can be encoded by the curvature, all at a classical level. 
 
  We proceed as follows. 
Given spatial gauge potentials $\mathbf{A}^1, \mathbf{A}^2$ on the spatial surface $\Sigma$, define  $\mathbf{C}:=\mathbf{A}^1-\mathbf{A}^2$ where $\mathbf{C}$ is a 1-form on $\Sigma$. We want to find out whether non-gauge related potentials can produce the same curvature tensor. Thus suppose $\mathbf{A}^1, \mathbf{A}^2$ are such that
\be \d \mathbf{A}^1=:\mathbf{B}^1=\mathbf{B}^2:=\d \mathbf{A}^2,\ee and so 
\be\d \mathbf{C}:=\d \mathbf{A}^1-\d \mathbf{A}^2=0.\ee  Now, if there are  $\mathbf{C}$ such that $\mathbf{C}\neq \d \lambda$ (for any $\lambda\in C^\infty(\Sigma)$), then $\mathbf{A}^1-\mathbf{A}^2\neq \d\lambda$ so $\mathbf{A}^1$ and $\mathbf{A}^2$ are not related by a gauge-transformation. That is, in this case, the two gauge potentials  are not in the same gauge-equivalence class, in spite of having the same curvature. Denoting equivalence classes with square brackets, we have $[\mathbf{A}^1]\neq [\mathbf{A}^2]$, while $\mathbf{B}^1=\mathbf{B}^2$.

By definition, such a $\mathbf{C}$ (with $\d \mathbf{C}=0$ and such that $\mathbf{C}\neq \d \lambda$) would be a member of $H^1(\Sigma):=\mathrm{Ker}\,\d^1/\mathrm{Im}\, \d^0\subset \Lambda^1(\Sigma)$, where $\d^1$ is the exterior derivative operator acting on the space of 1-forms on $\Sigma$, $\Lambda^1(\Sigma)$, and $\d^0$ is that same operator acting on smooth functions (or 0-forms). This space is called \emph{the first de Rham cohomology} of $\Sigma$,  and it is non-trivial only if there are loops in $\Sigma$ that are not contractible to a point: a topological condition. For such $\Sigma$, we can therefore find distinct equivalence classes $[\mathbf{A}^1]\neq [\mathbf{A}^2]$ that can nonetheless correspond to the same electric and magnetic field. 
(See  \citep{nounou2003ab} for a defense of the \lq{}topological\rq{} explanation of the A-B effect). 
This classical description of the origin of the effect also shows that the magnetic field does not need to vanish outside the solenoid. 

The non-trivial topology encoded in $H^1(\Sigma)$ is what \emph{allows} more than one  gauge equivalence class $[\mathbf{A}]$ on $\Sigma$ with vanishing magnetic fields. However, the particular class $[\mathbf{A}]$ is not itself encoded in the non-trivial topology.  In other words, the non-trivial topology of the region is an idealisation supposed to capture the presence of the infinite solenoid.   The particular shift of the interference fringes depends specifically on the content of this idealisation, i.e. on the current going through the solenoid, which corresponds to different equivalence classes of $[\mathbf{A}]$.\footnote{ Recently, \citet{ShechAB} and \citet{Earman2019} have challenged the idealisations associated with the Aharonov-Bohm effect, and \citet{DoughertyAB} has defended them. I stand with Dougherty.}

Thus suppose that $\Sigma=\RR^3-C$, where $C$ is a cylinder corresponding to the region occupied by the solenoid, where we could have two different field strengths, $\mathbf{B}$ and $\mathbf{B}\rq{}$. While elements of different equivalence classes  $[\mathbf{A}]$ and $[\mathbf{A}\rq{}]$ could be instantiated on $\Sigma$, they can only be extended to   $\RR^3$ for the matching field-strengths in $C$. The compatibility condition is given by Stoke\rq{}s theorem: 
\be\label{eq:flux_A-B}\left(i\oint_{\gamma_2\circ\gamma_1}\mathbf{A}\right)=\int_D\d \mathbf{A}=\int_D \mathbf{B}=\int_{D\cap C} \mathbf{B}, 
\ee
where $D$ is any two-dimensional disk bounded by $\gamma_2\circ\gamma_1$, and where we assumed the field strength vanishes only inside $C$.

A different set of worries concerns the possibility of a faster-than-light causal effect between the field inside the solenoid and the electrons causing the interference fringes. But one should be  careful in inferring anything about the time-dependent from the time-independent case:  the former necessarily involves the full set of dynamical Maxwell equations. In a tractable domain of operation of the solenoid,  once one takes into account both the electric and the magnetic A-B effects, it can be shown that there is no faster-than-light effect \citep{VanKampen1984}. Moreover, the time-dependent case admits a treatment similar to the one using homotopies, as above (see \citep{Gaveau2011}).

%In other words, $[\mathbf{A}]$ carries a local physical component---expressed in the magnetic field, or in the spatial part of the field-strength tensor, $\mathbf{F}$---and a topological one: expressed in the cohomological content of $\mathbf{C}$.  %That is,  different equivalence classes of the gauge potential may differ only insofar as they have different components in $H^1(M)$. %Both local and non-local components  are gauge-invariant.  %In other words, the equivalence classes of the gauge potential can be seen as a field which carries both local and non-local gauge invariant information.  %The  difference between the two gauge-equivalence classes, according to theory, supervenes on a non-local feature of the system. 

Now we can see that, in the treatment above, no quantisation is ever explicitly invoked. As we just saw, the A-B phase can be described as a purely classical physical property of the electromagnetic field. The entire discussion involved only classical electromagnetism and classical geometry; it just so happens that in order to \emph{experimentally} probe this property we need a superposition of an electron state along two different arms of an interferometer. 

Here we find a point of disagreement between me and \cite{Wallace_deflating}\rq{}s treatment of the A-B effect: while he admits that the effect involves only a classical background of electromagnetic fields, he also takes some quantum properties of matter to be essential to its physical significance. If one makes this alternative assumption, Wallace argues that, due to the difficulties in neglecting the backreactions of the quantum matter on the classical electromagnetic field, a consistent description of the effect could only be given within a full quantum-field-theoretic (QFT) treatment. But that picture also has its complications, for instance:  describing the local interactions between the electrodynamical field and the quantum matter in a gauge-invariant way; the need to invoke ever more complicated effects, such as vacuum polarization; etc.
  
  Why does Wallace escalate an explanation of the A-B effect all the way up to QFT?  He reasons as follows. After listing the merits of the holonomy formalism of electromagnetism,\footnote{In electromagnetism, the holonomy formalism takes operators $H_\gamma$ from closed curves $\gamma$ to complex numbers $\mathbb{C}$ with prescribed composition rules for segments of curves,  $H_{\gamma_1\circ\gamma_2}=H_{\gamma_1}H_{\gamma_2}$. These are related to the standard variables by $H_\gamma(\mathbf{A})=\exp{\left(i\int_{\gamma}\mathbf{A}\right)}$. They are a relatively popular choice for what constitutes the fundamental ontology of the theory, cf. \citep{Healey_book} and \citep{Belot1998}. } he points out a main flaw that is often left unnoticed: the formalism doesn\rq{}t include charged matter. Upon trying to include matter, it becomes clear that a representation that is separately gauge-invariant for matter and for the electromagnetic gauge potential is wrongheaded. Indeed, one of  Wallace\rq{}s central claims is that one should think instead of the electromagnetic and the matter fields as one single, indivisible object.\footnote{As  \citet[p. 14]{Wallace_deflating} says: \lq\lq{} It is tempting to think that the question can be innocently rephrased as: what kind of ontologies for the electromagnetic field and for the matter field are compatible with the theory? Tempting, but mistaken—and this is one of the main points of the paper. Since the gauge transformation thoroughly mixes the two together, there is simply no justification—as long as we wish our ontology to depend only on gauge-independent features of the theory—for regarding the two mathematically-defined fields as representing two separate but interacting entities, rather than as (somewhat redundantly) representing aspects of a single entity.\rq\rq{}}
Accepting this conclusion requires us to include a nowhere-vanishing matter field. But that is not a high price to pay for what is now within reach: as Wallace shows, we can give a completely local account of the A-B effect---and indeed a gauge-invariant local representation of electromagnetism with matter!---via a physical realization of the unitary gauge \citep{Wallace_unitary}. In order for the unitary gauge to be accessible, the matter field needs to be nowhere vanishing: an assumption that can only be defended at the quantum level. From here, consistency leads us to treat everything within QFT and we are off to the explanatory ladder!

   Of course, I am not saying that deeper, more fundamental explanations involving QFT are not welcome. These explanations are welcome, but not necessary.\footnote{ That is, as long as such a QFT dissolution of the A-B effect is not be in clear conflict with our preferred classical treatment in the relevant limits. Otherwise, it would look like a case of being misled by essentially classical features of the classical problem. In the case of \cite{MarlettoAB}, while they stop short of a full quantum-field-theoretic treatment, they claim to solve the locality puzzle in the quantum domain by showing that the A-B phase is locally mediated  by the entanglement between the charge and the photons.  I can’t hope to give a comprehensive review of that paper here. Suffice it to say that I am suspicious of their use of Coulomb gauge: they employ it but go on to dismiss (ibid. p.2-3) its significance by saying that physical effects and quantities don’t depend on the gauge; that seems to me to be precisely the issue. And while they are right to say that the \lq\lq{}field energy variation due to the charge,
point by point along the charge’s path\rq\rq{} can be stated gauge-invariantly, its relationship to the phase is gauge-dependent. A full comparison of the explanations in the two domains must be left for future work. I thank an anonymous referee for pressing this point.}
   
    What  I \emph{am saying}, then, is that  a perfectly satisfactory explanation exists strictly at the level of classical electromagnetism. As \citet[p. 26]{Stachel1982} remarks: \lq\lq{}it should be noted
that the crucial point—that the phase factor is of
physical significance —would be true for any wave
field, whether that field is classical or quantum
mechanical.\rq\rq{} But I would go beyond: while, in order to test theories experimentally we must include matter, classical vacuum electromagnetism is a perfectly cogent theory, with its own observables and physical properties, irrespective of the inclusion of matter or even quantization. To make a gravitational analogy: we may well derive approximate Kepler\rq{}s laws from an appropriate general relativistic model, but we can also understand them solely within Newtonian gravitation.  Of course, like all other theories, classical vacuum electromagnetism is false, and is to be supplanted by more accurate theories (e.g. quantized). Nonetheless,   as forcefully argued by \cite{Belot1998}, we gain much understanding of Nature  by seeking to interpret each successful theory in its own terms: \lq\lq{}Each of these theories informs us about our world,
despite their profound divergence of opinion concerning ontology\rq\rq{} (ibid, 558).\footnote{Earlier on, \citet[p. 551]{Belot1998} writes: \lq\lq{}To the extent that such interpretative judgments place constraints on our beliefs about where the actual world might sit in the space of possible worlds, they are indeed judgments about our world. There is a clear sense, then, in which the interpretation of false theories teaches us about this world. Our beliefs about our world are reflected in our understanding of our false physical theories; so getting clear on the content of a false theory is one way to make explicit our beliefs about our world. Admittedly, this is a strange way to learn about the world. But it is also a fruitful one for us: in the absence of a true theory, our false theories provide much of our understanding of the structure of the world.\rq\rq{}}

   But at this point I have not yet provided the reader with a sufficient defence of my proposed dissolution of the puzzles posed by the A–B effect. This is the aim of the present paper. In what follows, I will show that the effect can be understood as a classical, local, gauge-invariant, and geometric phenomenon, and that its puzzling features arise only when we insist on describing gauge theory using apparatus—such as potentials and gauge transformations—that obscure its underlying structure. The key tool will be a reformulation of gauge theory using vector bundles and covariant derivatives, introduced later in Section \ref{sec:vector}. %But even before we turn to that formalism, the core insight—that the A–B effect is an instance of holonomy—can already be brought into view by analogy with parallel transport in spacetime.

\subsection{The non-Abelian case}\label{sec:non_Ab}

In our current best theories of physics, electromagnetism emerges from a more fundamental type of interaction, known as \emph{electroweak}, through a process called \emph{spontaneous symmetry breaking}. It is besides the point of this paper to explain this process or electroweak theory in any detail. What is important is that the symmetry group of electroweak theory is non-Abelian, as is that of the theory of strong interactions. We thus have  good reasons to investigate effects similar to the A-B effect in the non-Abelian case.

In the Abelian case studied in the previous Section we can cleanly describe the many-to-one relationship between equivalence classes of the gauge potential and the curvature. In the non-Abelian case, we cannot. On the positive side, we can find an explicit example of a kindred non-Abelian A-B effect that does not depend on topology or, equivalently, on the idealisations used in the Abelian case.\footnote{For a treatment of the non-Abelian Aharonov-Bohm effect in a non-trivial topology, see \citep{Horvathy1986}. }
 
 In the non-Abelian case, the gauge potential is a Lie-algebra valued one-form, $A_\mu^I$, whose relationship to the field-strength tensor is:
 \be\label{eq:F} \mathbf{F}=\d \mathbf{A}-[\mathbf{A},\mathbf{A}],\quad\text{in coordinates:}\quad F_{\mu\nu}^I=\partial_{[\mu}A_{\nu]}^I-[A_\mu, A_\nu]^I,\ee
with the square bracket in the subscripts denoting anti-symmetrization, where $\nabla_\mu$ is the Levi-Civita covariant derivative on spacetime. For
 a Lie-algebra valued function $\xi:=\xi^I \epsilon_I:U\rightarrow \mathfrak{g}$, with coefficients $\xi^I\in C^\infty(U)$ the potential transforms as:
\be\label{eq:gauge_trans}\delta_\xi \mathbf{A}:=\d \xi+[\mathbf{A},\xi]=\D\xi, \quad\text{in coordinates:}\quad \delta_\xi A_\mu^I:= \pp_\mu\xi^I+[ A_\mu, \xi]^I=\D_\mu\xi^I,
\ee
where $\D_\mu(\bullet)=\pp_\mu(\bullet)+[A_\mu, \bullet]$, the gauge-covariant derivative. Thus: 
\be \delta_\xi F_{\mu\nu}^I=[\xi, F_{\mu\nu}]^I.
\ee

In this non-Abelian vacuum Yang-Mills case, we can find two non-gauge-equivalent potentials \(\mathbf{A}\) and \(\mathbf{A} \) with the same field-strength \(\mathbf{F}\neq 0\) on a simply-connected region.  A simple example is the following: take the gauge group \(SU(2)\) and base manifold \(\mathbb{R}^2\).
 The Pauli matrices, denoted as \(\sigma_1\), \(\sigma_2\), and \(\sigma_3\), form a basis for the Lie algebra $su(2)$:
\[
\sigma_1 = \begin{bmatrix}
0 & 1 \\
1 & 0
\end{bmatrix},
\quad
\sigma_2 = \begin{bmatrix}
0 & -i \\
i & 0
\end{bmatrix},
\quad
\sigma_3 = \begin{bmatrix}
1 & 0 \\
0 & -1
\end{bmatrix}.
\]
The Pauli matrices satisfy the following algebraic relations, known as the Pauli algebra:
\be\label{eq:Pauli}
\begin{aligned}
    \sigma_1^2 &= \sigma_2^2 = \sigma_3^2 = \mathsf{Id}, \quad \text{(where \(\mathsf{Id}\) is the identity matrix)} \\
    \sigma_i \sigma_j &= -\sigma_j \sigma_i \quad \text{for } i \neq j, \quad \text{(antisymmetry)} \\
    \sigma_i \sigma_j &= \delta_{ij}\mathsf{Id} + i \epsilon_{ijk}\sigma_k, \quad \text{(where } \epsilon_{ijk} \text{ is the Levi-Civita symbol)}
\end{aligned}
\ee

 Now consider
\be\label{eq:Asigma} \mathbf{A}_1 =  -i  \sigma_3 y dx + i  \sigma_3 x dy \ee
\be \mathbf{A}_2 = i \sigma_1 dx - i \sigma_2 dy \ee

We have 
\be\d \mathbf{A}_2=0= \mathbf{A}_1 \wedge \mathbf{A}_1, \quad \d \mathbf{A}_1=2i \sigma_3 dx \wedge dy=\mathbf{A}_2 \wedge \mathbf{A}_2.
\ee 
For  $\mathbf{F}= d\mathbf{A} + \mathbf{A} \wedge \mathbf{A}$ we get:
\be \mathbf{F}_1 =\mathbf{F}_2= 2i \sigma_3 dx \wedge dy.\ee
 If $\mathbf{A}_1$ and $\mathbf{A}_2$ were gauge-related, there should exist \(g \in C^\infty(\mathbb{R}^2, SU(2))\) such that \(\mathbf{A}_2 = g\mathbf{A}_1g^{-1} - dg\,g^{-1}\), in which case \(\mathbf{F}_2 = g \mathbf{F}_1 g^{-1}=\mathbf{F}_1\). That is,  \(\mathbf{F}_1\) should be invariant under such a \(g\), or, infinitesimally, $\mathbf{F}_1$ should commute with the generator of the transformation. Since \(\mathbf{F}_1 \propto \sigma_3\), and, from \eqref{eq:Pauli}, the only transformations that commute with $\sigma_3$ are generated by $\sigma_3$, we would have \(g_o = e^{i\theta\sigma_3}\) for some \(\theta(x, y)\). From \eqref{eq:Asigma}, since $\mathbf{A}_1$ only contains $\sigma_3$, we get that $g_o\mathbf{A}_1g_o^{-1} =\mathbf{A}_1$ and thus
\be g_o\mathbf{A}_1g_o^{-1} -dg_o\, g_o^{-1} = \mathbf{A}_1 - i d\theta \sigma_3.\ee  
Clearly, since this expression still only contains $\sigma_3$,  there is no \(\theta\) that can transform it into \(\mathbf{A}_2\). 

This concludes the example, and shows that in the non-Abelian case, we can have non-gauge-equivalent gauge potentials with the same field-strength, even in a topologically trivial background. This further supports the claim that the topological explanations invoked in the Abelian case are not essential to the phenomenon itself, even if non-trivial topology is necessary for that case. The deeper point is that the A–B effect arises whenever the equivalence classes of the gauge potential distinguish physical situations that the field-strength tensor does not. That feature will find its clearest mathematical expression once we shift to the vector bundle point of view, introduced   in Section \ref{sec:vector}.

\section{Spacetime analogies}\label{sec:grav_analog}

The puzzles posed by the A-B effect are not confined to electromagnetism: they reappear in the treatment of every fundamental interaction. In the previous Section, I described how  similar ideas apply to the strong and weak nuclear forces, and here we will see how they also apply  to gravity.  I will look at two gravitational analogies to the A-B effect: one using proper time (Section \ref{sec:proper}) and  one  using the parallel transport of vectors (Section \ref{sec:PT}). %Then, in Section \ref{sec:vector} I will argue that the second analogy is ideal for the discussions surrounding the A-B effect. 

\subsection{The local accrual of phase: a (dis)analogy with proper time}\label{sec:proper}
Thus far our investigation into the A-B effect has left us with only one puzzle: for each choice of gauge potential the profile of the phase gained along the trajectory will look different, thus there can be no physically significant local accrual of a phase   \citep[Section 6]{Healey2004}, \citep[Ch. 2]{Healey_book}. Indeed, by the remarks of Section \ref{sec:AB_EM}, were we to consider only a sub-region of $\Sigma$ that is simply connected, the vanishing of the field-strength on that sub-region implies the gauge potential could be also set to vanish there. One is led to suspect that the situation involves some kind of holism or non-locality. 
 
 \cite{Jacobs_PFB} tries to deflate this puzzle with a gravitational analogy that employs proper times:
 \begin{quote}
The following analogy is helpful. Consider the Twin Paradox [...] Just as we are interested in where the phase difference occurs in the Aharonov-Bohm effect, we may wonder when the age difference between the twins comes into existence. This is easy enough to answer with respect to certain planes of simultaneity. %[But] one can choose a convention such that the rocket-twin ages at the same rate as her earthbound twin on the outbound leg of her trip but then ages much more slowly on the return leg. 
[...] This result is analogous to the fact that one can choose a gauge such that $\mathbf{A}$ is zero over any open path. 
[...]

This implies that effects such as the Twin Paradox and the Aharonov-Bohm effect are holistic in this sense: although the total effect size (the age difference or interference shift) is measurable, there is no fact of the matter as to how this effect comes about as the result of small local differences. The final age difference between the twins is not the result of many small age differences that accrue locally. Likewise, the final phase difference in the Aharonov-Bohm effect is not the sum of the phase differences over infinitesimal paths. %Although puzzling, such holism is simply a consequence of the fact that value spaces are localised. 
 \end{quote}

 I am in almost complete agreement. But I must warn against a possible misconstrual of  the analogy between the  phase shift in the A-B effect and the lapse between the proper times of twins.  
 The proper times accumulate continuously, even if the lapse between them (before reunion) depends on a simultaneity surface. 
Looking more closely, we see that this lapse does not actually require closed loops: it only requires the selection of particular points along each trajectory (cf. \citep[Section 6]{Healey2004}). Given points along the trajectories, the proper times come along for the ride; there is no available choice to be made that would make the proper times vanish along any finite segment of the trajectories.  So there is a clear sense in which this effect, the lapse in proper times, \emph{is} incrementally accrued. (See \citep[Section 5]{Weatherall2016_YMGR} for a criticism that is similar to mine). 

I will argue this is an important difference with the gauge case.
    In the case of the electron\rq{}s incremental phases, we can\rq{}t  say where  they were picked up at all.     
    %The holonomy of spacetime vectors provide a closer analogy: any rotation of a parallel transported vector is relative to a fixed coordinate or frame basis, and yet we can talk about the disagreement of two such transported vectors upon the reunion of the paths.
 As we will now see, a better analogy between the gravitational and the gauge A-B effects requires the use of spacetime vectors.  I now turn to this. 

\subsection{The A-B effect for the parallel transport  of spacetime vectors}\label{sec:PT}	
There are by now several treatments of the analogues of the A-B effect within general relativity (cf. \cite{Dowker1967, Anandan1977, Ford_1981, Stachel1982}).
 The treatment here is closest in spirit to \citep{Dowker1967, Ford_1981, Stachel1982}. But unlike those papers, I will not exhibit  solutions of the Einstein equations or investigate the most general family of spacetimes that can incorporate the essential features of the A-B effect; and unlike \citep{Anandan1977} I am also not interested in the experimental setup required to verify this effect.  Thus, first, in Section \ref{sec:simple}, I will describe the simple setting that already exhibits the important elements of the A-B effect; then in Section \ref{sec:solder} I will describe and dismiss a well-known objection to the analogy between the gravitational and the electromagnetic A-B effects. 

\subsubsection{A simple realisation of the A-B effect via deficit angles}\label{sec:simple}
 As in the previous Abelian and non-Abelian cases, we would like to show that there is more to the covariant derivative than the curvature can distinguish. So  we take a closed curve $\gamma_1\circ \gamma_2$, a vector $v$ at the origin of $\gamma_1$ to be parallel transported,  and we fix the Riemann curvature tensor $R_{abcd}$ along  $\gamma_1\circ \gamma_2$. Can we still find sufficiently different covariant derivatives for which the total holonomy of $v$ (as it is parallel transported around $\gamma_1\circ \gamma_2$)  doesn\rq{}t vanish?  
 
 \cite{Stachel1982} pursues this question to some degree of generality for stationary metrics, and shows that the answer is in the affirmative.\footnote{He also shows how to couple a realistic physical system to the geometry in order to exhibit the effect.} But a simpler and more direct analog of the standard effect described in Section \ref{sec:setup} would present physically distinct situations in which the curvature $R_{abcd}$ remains zero in the entire region declared `accessible' to the system under investigation (a situation also described, even if briefly, in  \citep{Stachel1982}).  
  To realize this picture, in a  two-dimensional setting, take $\nabla$ as the Levi-Civita covariant derivative of a metric $g_{ab}$, and $\gamma_1$ and $\gamma_2$ to be (non-geodesic, of course) half-circles joining into a circle. Then: 
\\\noindent (1) consider  the parallel propagation of $v\in T_{\gamma_1(0)}M$ along $\gamma_1$ and $\gamma_2$, in Euclidean or Minkowski space; and 
\\\noindent (2) at the point at the center of the circle $\gamma_1\circ \gamma_2$  `cut out' a  wedge from the spacetime, encompassing an angle $\theta$, and then stitch spacetime back together along the edges of the wedge. This second situation creates a cone, with a singular curvature at its apex, whose value depends on $\theta$ (see figure \ref{fig2}). We now parallel propagate the vectors in the same manner as in (1).\footnote{ The same type of curvature defect could be obtained more realistically by a `cosmic string', for which we can then have the paths be geodesics; cf. \citep{Ford_1981}, and one can similarly obtain negative curvature by adding an angle to a cut.}
\begin{figure}[h!]
\center
\includegraphics[width=0.4\textwidth]{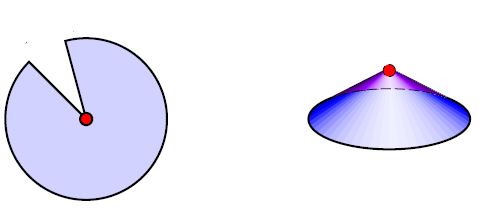}
\caption{Parallel transport of a spacetime vector along a closed curve enclosing a conical singularity. Although the curvature vanishes everywhere away from the apex, the holonomy around the loop is non-trivial. This illustrates how global features of parallel transport need not be reducible to local curvature—even in purely classical and local geometric settings.}\label{fig2}
\end{figure}

 In the first, but not the second situation, any vector will come back to itself, unrotated. In the second situation, there will be a relative rotation, depending on $\theta$.\footnote{To see this, picture a vector at angle $\alpha$ with respect to the rightmost edge. Upon identification of this edge to its left counterpart, that vector will still have angle $\alpha$ but now with respect to the left edge. Since the left edge is at angle $\theta$ with respect to the right edge, the overall rotation of the vector as it goes around a loop must be $\theta$.} Note also, that this difference cannot be attributed solely to a difference in path length: the same angle arises from the parallel transport around loops of arbitrary length around the apex of the cone. 

Experimentally, the singular curvature between the paths would affect the interference properties of a coherent beam of
particles such as neutrons, or  indeed, of any system whose state has a vector component, e.g. an axis of rotation of a gyroscope.  %Thus, for example, in situations like (ii), neutrons which traverse a region of space where the curvature is identically zero are nonetheless capable of detecting the effects of curvature in a far away region of space-time, i.e. at the conical singularities.

Once again, when we talk about excising the point corresponding to  the conical singularity from the spacetime, this is to be understood as an idealisation replacing the source of curvature.  So in this simple case, we can also understand this analogue to the A-B effect via cohomology, for there is a straightforward extension of the usual de Rham cohomology to flat vector bundles (see e.g. \cite[Ch. 5]{voisin}).  But the underlying point is that the Levi-Civita covariant derivative in the two situations---with and without a conical defect---will globally differ, even if the curvature doesn\rq{}t.\footnote{Note also that for simply-connected spacetimes, a vanishing curvature indeed means the metric is (isometric to) Minkowski spacetime. The question  whether, as in the more analogous non-Abelian case surveyed in Section \ref{sec:non_Ab}, we could get different covariant derivatives for the same non-zero curvature, is partially examined in \citep{Stachel1982}.} %However, as also discussed in Section \ref{sec:non_Ab}, in the case of non-vanishing curvature, the parallel transport will likely bear more path dependence, and so there could be no discernible A-B shift to speak of.}

Here is the important point about locality: the vectors are being parallel transported with respect to the covariant derivative, and that acts locally and smoothly (the discontinuity on the right-hand-side of the picture is only necessary if we wish to embed the geometry into $\RR^2$). The condition for the existence of this different covariant derivative is the non-trivial topology of space, idealised as a confined source of curvature. But the vectors being parallel transported don\rq{}t need to \lq{}feel\rq{} the distant properties of the curvature: all they know is the local covariant derivative.

Lastly, note that there \emph{is} a fact of the matter about how much rotation $v$ undergoes as it is parallel transported along each of the paths $\gamma_1$ and $\gamma_2$. The answer is \lq{}no rotation\rq{}! That must be the case since, under the standard interpretation, parallel transport defines a standard of \lq{}no rotation\rq{}.  When the student of differential geometry first encounters  the covariant derivative and parallel transport, what takes some getting used to is precisely this idea: viz. that even if we take $v$  along $\gamma_1$ without rotation, and  $v$ along $\gamma_2^{-1}$ without rotation, the two resulting vectors at the end of these paths may be rotated with respect to each other. Was this relative rotation continuously accrued along the paths? No. Where did this relative rotation take place? Nowhere. The parallel transported vectors simply disagree when they reach their common destination, but there is no build-up of that disagreement. The only mildly puzzling property of spacetime geometry that is illustrated by the setup above is the multiplicity of covariant derivatives that give rise to the same local curvature. Lastly, note that even jointly, the flat curvature and the non-trivial topology don\rq{}t give rise to a unique covariant derivative. We could, for instance merely cut out the same vertex, but with $\theta=0$: this would create a non-trivial topology, but there would be no associated shift in the parallel transport. 

%Indeed, in the next section, I will argue that the gauge A-B effect, in either the non-Abelian or Abelian case, can be understood \emph{precisely} as due to the parallel transport of vectors in vector bundles, and so can be interpreted  precisely  as in this spacetime example. First, we must describe and dismiss a possible source of disanalogy. 

    \subsubsection{A disanalogy due to soldering?}\label{sec:solder}
  A straightforward objection to collapsing the distinction between a gravitational A-B effect and the gauge A-B effect in the way I\rq{}m proposing is found in \citep{Anandan_1993}  (see also \citet[Secs. 6-7]{Healey2004}). The objection focuses on one sense in which  the spacetime vector rotation \emph{can be} construed as locally accrued, unlike in the gauge case. 
  
  Namely, unlike internal vectors,   tangent vectors are `soldered onto' spacetime, meaning that they are the tangents to curves in spacetime. This allows us to represent  the angle between the parallel transported vector and the tangent to the curve as locally accrued.%Thus, for example, if the vector was just tangent to one of the trajectories, the accrued difference would be a function of the total intrinsic acceleration (and thus related to the  difference in total elapsed proper times).
%\footnote{I take the  angle between the parallel transported vector and the tangent to the curve to be locally accrued even if that angle vanishes. In this point I depart from \citet[Section 5]{Weatherall2016_YMGR}'s kindred criticism (of Healey's argument for a disanalogy). For Weatherall argues  that a reference of constancy is only meaningful if the particle paths are geodesic, which would constrain the angle to vanish throughout motion. Thus, he says, there would be no local accrual.  But there is a difference between local accrual not being well-defined and a vanishing accrual. Moreover,  I also believe one could have a meaningful notion of constancy even for non-geodesics, e.g. for an accelerating rocket.} 
    
     But this  objection cuts ice only in the simplified two-dimensional treatment above; it dissolves when more detail is added. That is,  to assess the relative rotation of  the spin of a particle such as that of the neutron, i.e. to assess the relative rotation of polarization vectors, we must use Fermi-Walker transport. In other words, we are calculating a type of \emph{Thomas precession}, which is about the rotation of a spatial vector (a 3-vector),  which includes the rotation in the plane orthogonal to the timelike trajectory of the particle.  Of course, the angle between a polarization 3-vector and the tangent to the curve is also constantly zero. To put it differently: comparing the parallel transported 4-vector to the tangent to the trajectory allows us to locally determine the evolution of one degree of freedom of the 4-vector; but this still leaves open how the remaining polarization degrees of freedom---three if the particle is massive, two if it is massless---evolve along the trajectory. For these, all we can do in a coordinate independent way is what we had done before: compare a relative rotation upon the reconvergence of the paths.

  In the next section, I will argue that the gauge A–B effect—whether in the non-Abelian or Abelian case—can be understood precisely as due to the parallel transport of vectors in a vector bundle. In this setting, the analogy with the gravitational example above is not merely heuristic but exact. The formalism of vector bundles and covariant derivatives provides a natural framework for capturing this structural parallel, and it allows us to state the core claims about locality and classicality without appealing to gauge potentials, holonomies of internal spaces, or preferred gauges. This geometric framework will now be introduced.

 \section{The A-B effect without the gauge potential}\label{sec:vector}
 
 This is the central section of the paper. We have seen that versions of the Aharonov–Bohm effect arise across many fundamental interactions, and that the underlying phenomenon resists explanation in terms of curvature or topology alone. I now argue that this is no accident: the true core of the effect is simply parallel transport—nothing more.

To show this, I need to have a geometrical interpretation of the particle interactions on a par with the geometrical interpretation of general relativity. Thus I introduce a geometric formulation of gauge theory that avoids gauge potentials, principal bundles, and explicit reference to gauge symmetries. I call that the \emph{vector bundle point of view} of gauge theory. On this view, the gauge potential is analogous to a `coordinate expression\rq{} of an affine covariant derivative on a vector bundle, just as Christoffel symbols are the coordinate expression of the Levi-Civita covariant derivative in differential geometry. The covariant derivative is the infinitesimal version of parallel transport on these vector bundles; once we dispense with the gauge and work directly with covariant derivatives on vector bundles, the A–B effect takes on a strikingly familiar character: it is an instance of holonomy, precisely analogous to the parallel transport of spacetime vectors discussed in the previous section.
The effect arises naturally within classical differential geometry—so long as we use the right tools.

In Section \ref{sec:PT_gauge},  I introduce the technical machinery of vector and principal fiber bundles, and describe their relationship and use in gauge theory. In Section \ref{sec:vb_A-B} I will argue that we don\rq{}t need to have this detour through principal connections and principal bundles; we can talk about all gauge theories using only vector bundles and tensor products thereof. Although this marks a departure from the more familiar presentation of gauge theory in terms of gauge potentials and principal fiber bundles, I will briefly explain the motivation behind that traditional formalism, and then show why it is dispensable in the applications of interest here. Then, I will describe the A-B effect in a local, coordinate independent---or gauge-independent---form, using only parallel transport and covariant derivatives on vector bundles. The vector bundle formulation not only aligns more closely with the formal treatment of holonomy in general relativity, but also provides a conceptually cleaner account of the A–B effect itself. 
 %Here, I will apply this insight to more general vector bundles, not just those associated with the tangent and cotangent spaces of spacetime (vector and principal fiber bundles are concisely summarised in the appendix). In this way, I will argue that the A-B effect is solely about the disagreement of parallel transported vectors. %Any comparison of these vectors at mid-points of $\gamma_1, \gamma_2$ do depend on coordinates, as they do in standard differential geometry. An absolute comparison exists only when they are elements of the same vector space (or fiber, in the language of fiber bundles). 

 \subsection{Covariant derivatives with and without the gauge potential}\label{sec:PT_gauge}
 Gauge potentials are usually introduced as spacetime representations of an object---the principal connection form, $\omega$---that lives not in spacetime but on the principal fiber bundle (PFB). The use of PFBs seems necessary for gauge theory because different particle fields, such as electrons and quarks, live in different vector bundles, but particles that are charged under the same force must be parallel transported in step, e.g. they must feel the same field-strength of the connection. So there must be some enforced relationship between the covariant derivatives of different vector bundles.  PFBs provide a way to coordinate all of these covariant derivatives by defining the vector bundles on which the particles live as \emph{associated vector bundles} (cf. \citep{Weatherall2016_YMGR, Gomes_elements} for  expositions of this idea). Under this definition, a principal connection $\omega$ on the PFB uniquely induces a covariant derivative in each associated vector bundle, thus ensuring that parallel transport proceeds in step.

 Let us now review this argument in more detail. In Section \ref{sec:vec_bun} I will first describe the intrinsic covariant derivative in terms of vector bundles, then its relationship with a connection-form; then in Section \ref{sec:PFB}, I will tie these connection forms to a principal connection on a principal bundle and with the standard gauge potential.

 \subsubsection{Vector bundles}\label{sec:vec_bun}
   \begin{defi}[Vector Bundle]
   A vector bundle  $(E, M, V)$ consists of: $E$ a smooth manifold that admits the action of a surjective projection $\pi_E:E\rightarrow M$ so that any point of the base space $M$ has a neighborhood, $U\subset M$, such that, for all proper subsets of $U$,  $E$ is locally of the form $\pi^{-1}(U)\simeq U\times V$, where $V$ is a vector space (e.g. $\RR^k$, or $\mathbb{C}^k$) which is linearly isomorphic to  $\pi^{-1}(x)$, for any $x\in M$. \label{def:VFB}\end{defi}
 Note that the isomorphism between $\pi^{-1}(U)$ and $U\times V$ is not unique,  which is why there is no canonical identification of elements of fibers over different points of spacetime. Each choice of isomorphism is called `a trivialization' of the bundle. 
 \begin{defi}[A section of $E$] A section of $E$ is a map $\kappa: M\rightarrow E$ such that $\pi_E\circ\kappa=\mathrm{Id}_M$.  We denote the space of smooth sections by $\kappa\in \Gamma(E)$.
\end{defi}  
\begin{figure}[h!]
\center
\includegraphics[width=0.5\textwidth]{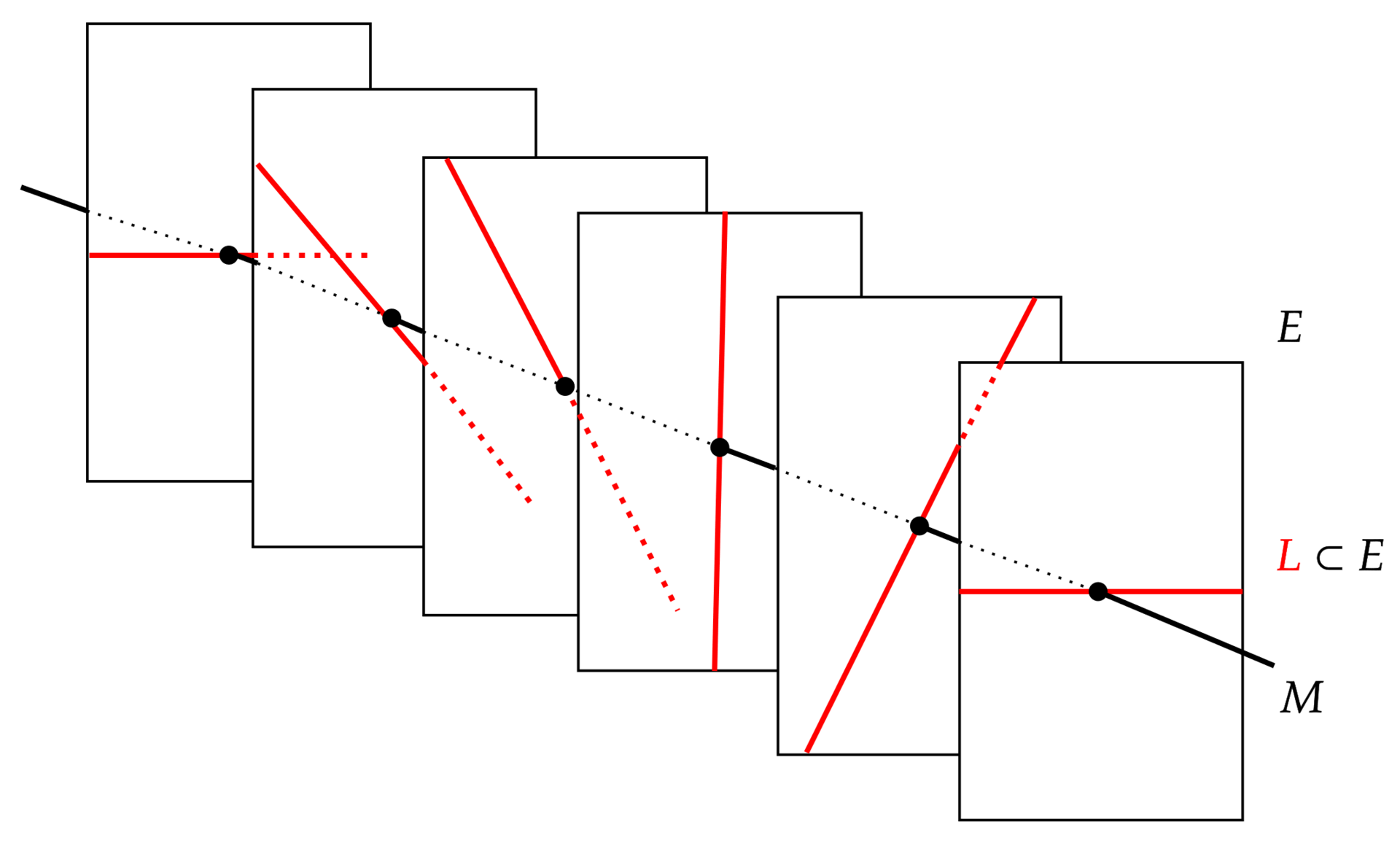}
\caption{A vector bundle with a two-dimensional fiber over a one-dimensional base space, with a section here called $L$. (Figure taken from Wikipedia)}\label{fig2}
\end{figure}
 Given a vector bundle $(E, M,V)$  a covariant derivative $\D$ is an operator:
\be
\D: \Gamma(E)\rightarrow \Gamma(T^*M\otimes E)\ee 
such that the product rule
\be\D(f \kappa)=\d f \otimes \kappa+f\D \kappa
\label{eq:affine_E}
\ee
is satisfied for all smooth, real (or complex)-valued functions $f\in \Gamma(M)$.

Thus we can define parallel transport  as follows:
   \begin{defi}[Parallel transport in a vector bundle]
Let $\D$ be a covariant derivative on $(E, M,V)$, $\mathbf{v}\in E_x$ and $\gamma(t)$ a curve in $M$ such that $\gamma(0)=x$. Then we define the parallel transport along $\gamma$ as the unique section $\mathbf{v}_h(t)$ of $E|_\gamma$ such that:
\be\label{eq:parallel} \D_{\gamma\rq{}}\mathbf{v}_h=0.
\ee
  \end{defi}
  The existence and uniqueness of this map is guaranteed for $\gamma\subset U$ some open subset of $M$, and it follows from properties of solutions of ordinary differential equations (cf. \citep[Ch. II.2]{kobayashivol1}). 

Here $\D$ is an operator, not a tensor. But by introducing a coordinate frame or basis, we can represent it as such. This is the same as for spacetime covariant derivatives, $\nabla$: it is only upon the introduction of a frame or basis that we find an explicit representation.

Let us see how this goes. Let the space of connections over $E$ be defined as:
\be \Delta(E):=\Gamma(T^*M\otimes \mathsf{End}(E)),
\ee
where $\mathsf{End}(E)$ are the linear, fiber-preserving endomorphisms of $E$,  isomorphic to $\Gamma(E^*\otimes E)$. So locally, much like Lie-algebra-valued one forms, these sections will take a vector on $M$ and spit out a linear transformation (i.e. a representation of $GL(V)$).

Now we must relate the covariant derivatives with the connections. 
Call $\mathcal{C}(E)$ the space of covariant derivatives for $E$. Given any $\D_o, \D\in \mathcal{C}(E)$, there exists a $\omega_{\D}\in \Delta(E)$ such that $\D_o-\D=\omega_{\D}$. Therefore, having fixed a choice of $\D_o$ the map: 
\begin{align}
\mathcal{C}(E)&\rightarrow\Delta(E)\nonumber\\
\D&\mapsto\D- \D_o\label{eq:omega}
\end{align}
is a bijection between the space of covariant derivatives and the space of connections. That is, the space of covariant derivatives is an affine space over the vector space of connections. In practice, we pick $\D_o$ via a trivialisation of $E$, i.e. we pick an isomorphism $\pi^{-1}(U)\rightarrow U\times V$, so that sections of $E_{|_U}$, i.e. elements of $\Gamma(E_{|_U})$, become functions: 
\be \kappa: U\rightarrow V.
\ee
 In such a trivialisation, we can take $\D_o\rightarrow \d$, which acts only on functions on $U$, not on their value on $V$. Thus, given a trivialisation and the subsequent identification of $\D_o$ with $\d$,  the connections parametrise the space of covariant derivatives. 
 It is only at this step---after a local trivialisation---that the covariant derivatives are described as 1-forms valued on $\mathsf{End}(E)$. %Thus, all representations of the covariant derivative as connections on spacetime require such a choice. 
 
  Now, with a trivialisation in place,  for some section of a general  real (or complex) vector bundle $\kappa\in \Gamma(E)$, we locally write $\kappa=\kappa^i\mathbf{e}_i$, and the covariant derivative  of $\kappa$ becomes: 
\be\D\kappa=d\kappa^i\otimes{\mathbf{e}_i}+\kappa^i\D \mathbf{e}_i.\ee

Summing up the construction thus far: an east way to relate a covariant derivative given in \eqref{eq:affine_E} explicitly to a connection is to  pick out a $\D_o$. We do that by picking a frame for $E$ (i.e. a local section $\sigma$ for the  bundle of linear frames over $E$, $L(E)$), call it $\{\mathbf{e}_i\in \Gamma(E_{|U}), i=1,...,k\}$ and abbreviate it by $\{\mathbf{e}_i\}$. Now we can represent the covariant derivative directly in terms of this frame.

 If $E$ is endowed with further structure, say, an inner product,  we usually  require the covariant derivative to preserve that structure, so that parallel transport is well-defined within the bundle.  Since $\D$ preserves the structure of the fiber,  it will preserve that structure of the frame $\{\mathbf{e}_i\}$ (e.g. being orthonormal). That means we can write, for $X\in(T_xM)$, 
\be \D_X\mathbf{e}_i=\omega^j_i(X)\mathbf{e}_j,\ee
where $\omega_i^j\in\Gamma(T^*U)$ are (matrices of) one-forms.  
We interpret these matrices of one-forms as follows: a linear transformation of $E_x$ is an element of $\mathsf{End}(E_x)$, and  $\mathsf{End}(E_x)\simeq E_x^*\otimes E_x$, so we can describe the extent to which the chosen basis is non-parallel along a certain direction by a 1-form valued on $E\otimes E^*$, which we write as:
\be\label{eq:omega_frame}\omega=\omega_i^j\otimes \mathbf{e}^i\otimes \mathbf{e}_j.\ee

Finally, the covariant derivative  of $\kappa$ becomes: 
\be\label{eq:D} \D\kappa=d\kappa^j\otimes{\mathbf{e}_j}+\kappa^i\omega_i^j\otimes\mathbf{e}_j.\ee
Note the double role of the frame: we use it to uniquely split $\D$ into $\d+\omega$, and then we use it to describe the particular coefficients of the (spin) connection $\omega$.

To define \(\Omega\) in terms of \(\D\), we proceed in the usual way: 
%given vector fields \(X, Y \in \Gamma(TM)\), the curvature is the bundle-endomorphism valued two-form \[\Omega(X,Y) := \D_X \D_Y - \D_Y \D_X - \D_{[X,Y]},\]
\begin{defi}[Curvature]\label{def:curv}
Given a covariant derivative $\D$ on a vector bundle $E$, the \emph{curvature tensor}  is the unique multilinear bundle map
\[
\Omega : TM \otimes TM \otimes E \to E \quad : \quad (X, Y, v) \mapsto \Omega(X,Y)\kappa
\]
such that for all $X, Y \in TM$ and $\kappa \in \Gamma(E)$,
\[
\Omega(X,Y)\kappa = \left( \D_X \D_Y - \D_Y \D_X - \D_{[X,Y]} \right)\kappa,
\]
where \([\cdot,\cdot]\) is the Lie bracket of spacetime vector fields.
\end{defi}
We can see the curvature then as an element of $\Omega : TM \otimes TM \otimes \mathsf{End}(E)$, i.e. as a map valued on the endomorphisms of $E$ (the fiber-linear transformations that are not necessarily automorphisms). Since $\mathsf{End}(E)\simeq E^*\otimes E$,  with a frame $\{e_i\}$ for $E$ we could write $\Omega=\Omega^i_je^j\otimes e_i$, where $\Omega^i_j$ are 2-forms, though we are in no obligation to do that. 

\subsubsection{Principal bundles}\label{sec:PFB}
In most physical theories, one has many vector bundles: one for each particle type, and thus many accompanying covariant derivatives. But different particle types are often charged under the same force, e.g. both electrons and quarks have electric charge. So there must be a sense in which their covariant derivatives march in step. A natural way to enforce this coincidence is to have the covariant derivatives of a family of vector bundles all be induced by a single principal connection. In the words of \citet[p. 40]{Jacobs_PFB}: 
\begin{quote} [I]t is only when we consider more than one field that the principal bundle becomes relevant. For if distinct matter fields couple to the same Yang-Mills field, it is useful to represent the latter  `by itself’ on a principal bundle. The claim that both matter fields couple to the same Yang-Mills field then translates into the fact that both vector bundles are associated to the same principal bundle. But it is a problem for this approach that the two fields survey the same connection as a matter of brute fact. There really are two connections: one defined over the first associated bundle, and one defined over the second. These connections are the same only in the sense that we can represent both with the same connection on a single principal bundle. 
\end{quote}

Here is how PFBs accomplish this in  more detail. Given a PFB $(P, M,G)$, a principal connection is a Lie-algebra-valued 1-form $\varpi:TP\rightarrow \mathfrak{g}$ (see \eqref{eq:omega_defs}).\footnote{It is pedagogic to understand the basic ideas for $(P,M,G)$ a bundle of frames (see Appendix), i.e. $p\in P$ is a frame over $x=\pi(p)$, and the gauge group changes the frame at each point. In this case the horizontal directions at $p$ can be understood as the infinitesimal parallel transport of that frame to frames over nearby spacetime points. } Such a connection defines a horizontal lift of a curve $\gamma\in M$ to a curve $\gamma_h$ in $P$ through $p\in \pi^{-1}(\gamma)$ as the unique curve going through $p$ such that, along its extension,  $\varpi(\gamma_h)=0$.  And given a horizontal lift, it is easy to define parallel transport in an associated bundle: given $\mathbf{v}=[p, v]\in E_x$, where $E=P\times_\rho V$ (see equation \ref{eq:AVB}), the parallel transport of $\mathbf{v}$ over $\gamma$ is given by $[\gamma_h, v]$.  This ensures  the marching in step of all the  covariant derivatives that a principal connection on $P$ defines on all the associated vector bundles.

But we still need to apply this formalism to fields or particle trajectories that traverse spacetime, and which cannot be embedded in $P$ without some choice or another. The standard way to obtain a representation of $\varpi$ on a spacetime region is to pick a section of $P$. I.e. for $U\subset M$ we define: 
 \begin{defi}[Local sections of $P$]  are maps $\sigma:U \to P$ such that $\pi\circ \sigma = \mathrm{id}$.\footnote{In the case of the bundle of frames $L(E)$, a section amounts to a choice of frames $\{\mathbf{e}_i\}$, where $\mathbf{e}_i\in E$ are linearly independent (and e.g. orthogonal, depending on whether $E$ has further structure). }
\end{defi}
A section functions as a \lq{}choice of coordinates\rq{} on the principal bundle:  a particular coordinate system that locally trivializes $P$, i.e. it realises an isomorphisms between $\pi^{-1}(U)$ and $U\times G$.

 The gauge potential that appears in the A-B effect, $\mathbf{A}$, is a spacetime representation of $\varpi$ that depends on a choice of $\sigma$. I.e. 
 \be \mathbf{A}=\sigma^*\varpi.
 \ee
   In possession of a section of a principal fiber bundle, we can identify $\omega^{i}_j$ appearing in \eqref{eq:D} with ${A}^i_j$, understood as a connection valued in a representation of the Lie algebra on the vector space that constitutes the typical fiber of the vector bundle. If the typical fiber is isomorphic to $\bb C$ (or $\RR$), then  $\omega_i^j$ is just a complex (resp. real) number, i.e. it is a complex (or real)-valued one form, like the vector potential $\mathbf{A}$. 
  
% The question remains: how can we represent the A-B effect without the gauge potential, or rather,  without some implicit choice of coordinates on $P$ or frames of $E$?  

 \subsection{All we need are vector bundles}\label{sec:vb_A-B}
 
After explaining the use of PFBs for coordinating covariant derivatives, Jacobs (ibid. p. 41) then goes on to discard the approach that he calls \lq{}deflationary\rq{}, in which: \begin{quote}Neither the principal bundle nor the [principal] connection on its own represent anything physical. Rather, it is the induced connection on the associated bundle that represents the Yang-Mills field. This approach has difficulties in accounting for distinct matter fields coupled to the same Yang-Mills field.\end{quote} The issue, as he sees it, is that
 \begin{quote}
  [O]n the deflationary approach there is no independent Yang-Mills field that the associated bundle connections supervene on. This makes it seem somewhat mysterious that these connections are equivalent. The coordination between associated bundles begs for a ‘common cause’ in the form of an independently existing Yang-Mills field.\footnote{Jacobs instead defends the \lq{}inflationary approach\rq{}, which: \lq\lq{}reifies not the principal bundle but the so-called ‘bundle of connections’. The inflationary approach is preferable because it can explain the way in which distinct matter fields couple to the same Yang-Mills field.\rq\rq{} As I have argued, it is not preferable in this sense. }
 \end{quote}

In \cite{Gomes_internal} I showed that this criticism can be overcome, and the deflationary approach rehabilitated. The introduction of PFBs is unnecessary if particles that interact are all sections of the same vector bundles or of tensor products of the same vector bundles. Tensor products over a vector bundle inherit the same covariant derivatives by construction. In this case, parallel transport of the vector bundles in question automatically march in step.  In this case we have at a hand a natural \lq{}common cause\rq{} for the coordination of covariant derivatives, without the introduction of principal bundles. I will here call this \emph{the vector bundle point of view} of gauge theory.

In more detail, given two  vector bundles, $E, E\rq{}$, a covariant derivative on $E$ will induce a covariant derivative on $E\rq{}$ whenever $E\rq{}$ is equal to a general tensor product involving $E$ and its algebraic dual, $E^*$. 
In more detail, given $E$ a vector bundle with covariant derivative $\D$, and $E^*$ its dual, we define, for sections $\kappa\in \Gamma(E)$ and $\xi\in \Gamma(E^*)$: 
\be  d(\langle \xi, \kappa \rangle)(X) = \langle \nabla_X^* \xi, \kappa \rangle + \langle \xi, \nabla_X \kappa \rangle,
\ee where here angle brackets represent contraction. The generalisation to  arbitrary tensor products is straightforward due to multilinearity.

In this view, there are no \lq{}gauge groups\rq{}; there are only groups of automorphisms of these vector bundles, $\mathsf{Aut}(E)\subset \mathsf{End}(E)$.  The distinction between Abelian and non-Abelian gauge theories then concerns these automorphisms. In particular, one-dimensional vector  bundles---such as those with typical fiber isomorphic to $\bb C$---give rise to Abelian groups of automorphisms. 

This formulation might seem at first sight insufficient for describing all the myriad gauge theories  used in physics. Indeed  some gauge theories, for instance those described by exceptional Lie groups,  lie outside of the scope of this interpretation.  But, as far as the standard model of particle physics goes,  all particles are represented as sections of associated bundles for  principal fiber bundles whose  structure groups are $SU(n)$ for some $n$.  But under any given representation of $SU(n)$, these associated bundles can be alternatively obtained as a tensor product of a fundamental vector bundle.  
 In cases such as these,  a covariant derivative on a single vector bundle represents one fundamental type of interaction. 
 
 This picture omits PFBs as well as spacetime representatives of principal connections.  
Gauge bosons, $A_\mu^I$, unlike the matter fields represented by fermions, are replaced by objects that are \emph{not}  sections of vector bundles. Here one should rather think of the physical content of gauge bosons as  structural features representing the geometry that guides the dynamics of the matter fields.  In other words, here  it is the geometry of the vector bundle encoded in the covariant derivative that is primary: (what are usually called) gauge bosons emerge from particular choices of representation of that covariant derivative, and so involve some redundancy.

 Now we would like to recover the familiar A-B effect from parallel transport around a loop, and for that we need to make a choice of trivialisation. First, using \eqref{eq:parallel} we   define an isomorphism: 
  \begin{align} \mathsf{PT}_\gamma(t): E_x\rightarrow E_{\gamma(t)}\nonumber\\
 \mathbf{v}\mapsto \mathbf{v}_h(t).\label{eq:PTmap}
  \end{align}
   Now, in a trivialisation over a subset $U\subset M$, i.e. given a choice of frame $\{\mathbf{e}_i\}$, we use \eqref{eq:D} and \eqref{eq:parallel} to write for a parallel transported vector:
  \be\label{eq:d_h} \d {v}_h^j\otimes{\mathbf{e}_j}+{v}_h^i\omega_i^j\otimes\mathbf{e}_j=0.\ee 
  We can solve \eqref{eq:d_h} to describe the relationship between the initial and final components of the parallel transported vector by using the path-ordered exponential. That is we obtain $\mathbf{v}_h$ at $\gamma(1)=y$ given its value at $\gamma(0)=x$:
  \be\label{eq:PTv} \mathbf{v}_h(y)=\left(P \exp \int_{x}^{y}\omega(\gamma\rq{})\right)\mathbf{v}_{h}(x).
  \ee
  In the Abelian case, for a closed loop, we get back the total phase shift of the A-B-effect, given in \eqref{eq:phase_A-B}. So, in general, we can talk about the total shift in the A-B-effect without the appearance of gauge dependence; it is nothing but the holonomy operator:
  \be\label{eq:PT} \mathsf{PT}_{\gamma_2\circ\gamma_1}\in \mathsf{Aut}(E_x),\ee
(since this map preserves structure, it is an automorphism of $E_x$).  

Of course, in order to extract a scalar value from this operator in the non-Abelian case, one must  apply to it a further scalar-valued operation, such as the trace, which assuming the typical fiber is endowed with an inner product, $\langle \cdot, \cdot\rangle$, can be written explicitly in terms of a frame as:
\be\label{eq:trace}\mathsf{Tr}(\mathsf{PT}_{\gamma_2\circ\gamma_1})=\sum_{i=1}^k  \langle \left(P \exp \int_{\gamma_2\circ\gamma_1}\omega(\gamma\rq{})\right) \mathbf{e}_i(x), \mathbf{e}_i(x)\rangle_x.\ee

  This concludes my presentation of the vector bundle point of view as an alternative to the use of principal and associated bundles. % and its resolution of the central puzzle about locality and gauge-invariance involved in the generalized A-B effect.  This resolution  avoids mention of the gauge potential and of gauge symmetry. Now I will defend this resolution from criticism by relying on the deep analogy to the holonomy of spacetime vectors.

\section{Criticisms}\label{sec:critic}

To recap, according to \eqref{eq:PT}, we can describe \emph{a generalised, classical A-B effect} for both spacetime and internal vectors and tensors. The puzzle posed by the generalised A-B effect relies on  a  difference between  two   situations described by two different covariant derivatives whose associated curvature is identical (or even identically zero) in the accessible regions. 

There are two possible objections to a deflation of the puzzle by the use of the vector bundle formalism. The first, is that I have not defended the formalism from the same pitfalls of the holonomy variables; namely, the non-locality and insufficient expressive resources for formulating the dynamics of the theory. I will discuss this in Section \ref{sec:dyn}.  The second objection, dealt with in Section \ref{sec:diffeos}, is orignal, but one I am preempting. It concerns the redundancy due to the active interpretation of background automorphisms of the vector bundle. 

\subsection{Dynamical variables}\label{sec:dyn}

This paper has focused on kinematics, or the geometry of a family of theories. The core technical proposal of this paper is that gauge theories should be formulated in terms of families of four-tuples \((E_i, M, V_i, \D_i)\), one for each fundamental interaction---e.g., in the Standard Model, \(i = 1, 2, 3\). Here, \(M\) is spacetime, \(E_i\) is a vector bundle with typical fibre \(V_i\), and \(\D_i\) is a covariant derivative acting on sections of \(E_i\). Matter fields charged under the \(i\)-th force are then represented as sections of \(E_i\) or of tensor products thereof. This yields an intrinsically geometric representation of the matter content of gauge theory. I have avoided important questions that lie in the transition between kinematics and dynamics, such as that of minimal coupling, which will not be treated in this paper. %As I will describe below, the gauge bosons, $A_{i\mu}^I$ are \emph{not} geometric in the same way as the matter fields. 

Although the focus was clearly kinematical, unlike more abstract approaches that attempt to dissolve the A--B puzzle---such as Healey's holonomy formalism \citep{Healey_book}---the vector bundle formulation still allows us to write down the equations of motion of Yang--Mills theory in a fully local and geometrically transparent way, using $\D_i$.  The entire formalism unfolds within standard differential geometry.

To be precise: in vacuum Yang--Mills theory on a four-dimensional spacetime, the field equations are usually written as:
\[
*\d_\D*\Omega = 0,
\]
where \(*\) is the spacetime Hodge operator  \(\Omega\) is the curvature two-form, and $\d_\D$ is the gauge-covariant exterior derivative. We need to translate these constructions into ones that make only reference to the vector bundle.  

Here we want a model of the vacuum theory for a single force, to be  specified by a four-tuple \((E, M, V, \D)\), with \(\D\) as the sole dynamical variable. Matter fields, as before, are sections of \(E\) or of tensor products thereof, and thus automatically couple to \(\D\) in the right way.

With respect to the curvature, $\Omega$, defined in Definition \ref{def:curv}, it is an $\mathsf{End}(E)$-valued map, but this doesn\rq{}t present further difficulties. The trace operation is defined as $\mathsf{Tr}: \mathsf{End}(E)\rightarrow C^\infty(M)$, and so can be included in a Lagrangian definition of the theory. Since $\mathsf{End}(E)$ is closed under composisiton, we can obtain a Lagrangian 4-form for the action: 
\be \mathcal{L}=\mathsf{Tr}(\Omega\wedge *\Omega).
\ee

As to the exterior covariant derivative, an $E$-valued 0-form is just a section of the bundle $E$. That is,
\[
\Omega^0(E) = \Gamma(E).
\]
Thus the covariant derivative is a linear map
$\D : \Omega^0(E) \to \Omega^1(E)$, which can be uniquely extended to an {exterior covariant derivative}
\[
d_\D : \Omega^r(E) \to \Omega^{r+1}(E), 
\]
defined by the Leibniz rule, which is specified on   tensors of the form $\lambda \otimes \kappa$ and extended linearly:
\[
d_\D(\lambda \otimes \kappa) = d\lambda \otimes \kappa + (-1)^{r} \lambda \wedge \D \kappa
\]
where $\lambda \in \Omega^r(M)$.%so that $\deg \lambda = r$, $s \in \Gamma(E)$ is a section, and $\omega \wedge \nabla s$ denotes the $(r+1)$-form with values in $E$ defined by wedging $\omega$ with the one-form part of $\nabla s$. Notice that for $E$-valued 0-forms, this recovers the normal Leibniz rule for the connection $\nabla$.

 In any case, we find a formulation of the laws where no gauge potential appears, and no reference is made to a principal bundle. The structure is local, geometric, and manifestly coordinate-free.
     
   \subsection{A remaining redundancy?}\label{sec:diffeos}  
  Let me now pre-empt a worry that makes a fleeting appearance in \citet[Sec.~4.2.2]{Healey_book}'s criticisms of \citet{Leeds1999}, concerning the role of active linear automorphisms in the vector bundle formalism.

As noted earlier, \citet{Leeds1999} articulates a view of gauge theory that shares important features with the one defended here—what \citet[p.~99]{Healey_book} describes as a ``hybrid between a principal fiber bundle and an associated vector bundle''. But Leeds restricts attention to the Abelian case, and a lack of mathematical precision leads to a fatal equivocation between the fibres of the principal bundle (isomorphic to $U(1)$) and those of the associated vector bundle (isomorphic to $\mathbb{C}$). Healey seizes on this confusion and identifies apparent contradictions, which he takes as grounds to dismiss Leeds' approach entirely. Since the publication of \citep{Healey_book}, this dismissal has largely gone unchallenged in the literature. For the mathematically rigorous formulation I have provided here, no such equivocation persists, no contradiction is found, and so these criticisms lose their bite.

     But, more importantly, in the course of formulating his argument, Healey also criticises the vector bundle formalism for having left the \emph{active} linear automorphisms as a remaining source of redundancy. This I  must still address.      
 It should be no surprise that  I\rq{}ll address it by  first noticing that precisely the same worry applies to the case of spacetime tensors.
   
   Recall the familiar distinction in general relativity between passive diffeomorphisms (coordinate changes) and active ones (which relabel points of the manifold). The Levi-Civita covariant derivative $\nabla$ is not invariant under active diffeomorphisms, but it transforms \emph{covariantly} as an operator—it is a geometric operator. The same holds for the covariant derivative $\D$ in the vector bundle formalism. 
   
 In more detail, we can formulate the corresponding active interpretation of gauge transformations by considering two fibre-wise linearly isomorphic vector bundles, $E, \widetilde{E}$, over $M$. 
   Two covariant derivatives in two linearly isomorphic vector bundles are equivalent if they are related by the conjugation by the linear isomorphism (here a diffeomorphism $f:E\rightarrow{\widetilde{E}}$ such that $\pi_{E}\circ{f}=\pi_{\widetilde{E}}$, where $f$ 
takes $\pi_{E}^{-1}(x)\rightarrow{\pi_{\widetilde{E}}^{-1}(x)}$ by a linear isomorphism). This relation guarantees that the following diagram commutes (for all $X\in\Gamma(TM)$):
\begin{eqnarray*}
\Gamma(E) & \xrightarrow{\D_X\phantom{1}} & \Gamma(E)\\ 
f\downarrow & \phantom{\xrightarrow{\D_X\phantom{1}}} & \downarrow{f}\\ 
\Gamma(\widetilde{E}) & \xrightarrow[\widetilde\D_X]{\phantom{\D_X{1}}} & \Gamma(\widetilde{E})\\
\end{eqnarray*} 
Thus we can represent the covariant derivative $\D$ under a bundle isomorphism obtaining a new covariant derivative. I.e. for all $X\in\Gamma(TM)$ and $\kappa\in \Gamma(E)$ :
\begin{equation}\label{conex}\D_X(\kappa)=f^{-1}\widetilde\D_X({f\kappa})\Rightarrow\widetilde\D_X=f\D_X{f^{-1}}\end{equation} 
The covariant derivative $\D$ transforms \lq{}tensorially\rq{} as a differential operator under automorphisms of the bundle (cf. \cite[Ch.~2]{Palais_book} for a description of local differential operators on vector bundles).  It is \eqref{conex} that allows us to see the covariant derivative \(\D\) as geometric. 

Nevertheless, we are left with isomorphic representations of a given vector bundle, including its covariant derivatives.  Should these isomorphic representations worry us? 

Here, the debate surrounding the ``hole argument'' in general relativity becomes directly relevant (see e.g. \citep{Pooley_Read, GomesButterfield_hole1} for recent reviews). The issue is whether isomorphic models represent distinct physical possibilities, and the current framework brings the same philosophical choices into play. One prominent view, often dubbed ``internal sophistication'' (see e.g. \citep{Dewar2017, Jacobs_Inv, Jacobs_PFB, Samediff_1a}), holds that redundancies arising from automorphisms of a fixed background structure do not correspond to physical differences. This principle applies in the vector bundle formulation of gauge theory just as it does in  the principal bundle formalism or in GR (see \cite{Jacobs_PFB} for a thorough defense of sophistication for the principal bundle formulation of gauge theory). 

There remains one potential reason to doubt the strength of the analogy between the Aharonov–Bohm effect in internal vector bundles and in the tangent bundle. In the standard (classical) A–B setup, the electron’s trajectories are typically treated as fixed. One might then think that, in the tangent bundle case, the presence of a soldering form fixes the identification between the fibres and the underlying manifold structure—so that once a curve is specified, parallel transport and therefore the holonomy are uniquely determined. If so, this would mark a disanalogy: in the tangent bundle, fixing a curve might eliminate any remaining gauge redundancy—active or passive—whereas in the internal case, no such identification exists to play an analogous role.
 
 But this is not so. Even if the curve $\gamma$ remains fixed, active diffeomorphisms that preserve $\gamma$ can act non-trivially on the vectors being transported along it.
The holonomy remains sensitive to such transformations, and the resulting ambiguity mirrors precisely the freedom present in the internal bundle case. 

Formally, consider the parallel transport map \eqref{eq:PTmap} with $E = TM$ and $\D \to \nabla$, a spacetime covariant derivative such as the Levi-Civita derivative. Even when $d: M \to M$ is a diffeomorphism that preserves the curve $\gamma$, it acts  non-trivially on the fibres along $\gamma$. For the covariant derivative still transforms:
\[
\widetilde\nabla_\gamma Y = d_* \big( \nabla_\gamma (d_*^{-1} Y) \big),
\]
for $Y \in \Gamma(TM|_\gamma)$. This is just the covariance condition \eqref{conex}, with diffeomorphisms in place of bundle automorphisms. Parallel transport also transforms covariantly;  for $\mathbf{v} \in T_x M$:
\be\label{eq:cov_PT}
\mathsf{PT}_\gamma(t)(\mathbf{v})= d_*^{-1} \big(\widetilde{\mathsf{PT}}_{(d(\gamma(t))}(d_*\mathbf{v})\big), 
\ee
where $\widetilde{\mathsf{PT}}$ is parallel transport according to $\widetilde\nabla$. Of course,  \eqref{eq:cov_PT} can still be non-trivial even when $d(\gamma(t))=\gamma(t)$.

Thus for both the spacetime tangent and internal bundle cases, parallel transport along a fixed curve varies under active automorphisms of the fixed background structure. Incidentally,  for one-dimensional vector bundles as in the case of electromagnetism, even though the covariant derivative varies as per \eqref{conex},  the holonomy itself---whose transformation has the same form as  \eqref{eq:cov_PT}, but with $d(\gamma(t))\rightarrow\gamma(t), \mathbf{v}\in E_x$ and $d\rightarrow f$---is already invariant; this is what we expected from Abelian theories.\footnote{But one can always extract invariant quantities for the other cases as well:  by taking the trace of the holonomy in both the tangent bundle and the higher dimensional internal vector bundles, as in \eqref{eq:trace}. Also note that we have here considered holonomy around a single closed curve, with $\mathsf{PT}_{\gamma_2 \circ \gamma_1} \in \mathsf{Aut}(E_x)$. The set of all such holonomies at a fixed base point $x$ defines a subgroup $\mathsf{Hol}_{(x)}(\D) \subseteq \mathsf{Aut}(E_x)$. On simply connected regions, $\mathsf{Hol}_{(x)}(\D)$ is conjugate across base points, so we may refer to the base-point-independent \emph{holonomy group} $\mathsf{Hol}(\D)$. Since isomorphic bundles share holonomy groups, this group is also an invariant.}

The upshot is this: whatever one’s stance on the hole argument, no one takes it to undermine the local geometric significance of holonomy in spacetime. Holonomy remains both well-defined and physically meaningful, despite varying under active diffeomorphisms. Why, then, should the situation be any different for general internal vector bundles? If there is a principled disanalogy, it has yet to be demonstrated—and the burden of proof lies squarely with those who claim it exists.

\subsection{The case for the vector bundle point of view}

Let me be clear from the start: the vector bundle point of view defended here is \emph{not mathematically superior} to  any other view of gauge theory, e.g. via principal bundles. At the end of the day, mathematical equivalence prevails. My clain is that this point of view offers a conceptually cleaner framework for addressing questions of gauge invariance and locality—not only for the specific case of the electromagnetic A–B effect, but also for more general Yang-Mills theories. I judge there to be two core advantages over the familiar principal bundle formulation and its cousins: (a)  covariant derivatives are  coordinated across interacting matter fields without appeal to auxiliary structures other than tensor products; and (b) we obtain an explicitly local, geometric characterisation of the A–B effect.

\subsubsection{Coordination}
Let us start with (a). Consider the status of $\D$: it is simply a covariant derivative on a vector bundle—a thoroughly ordinary geometric operator. And yet, in the literature, it is rarely deployed to clarify conceptual questions about locality and gauge invariance. Why is that? %Why, for instance, has it not been more widely recognised that substituting $A_\mu$ with $\D$ in electromagnetism yields a fully gauge-invariant formulation of the vacuum theory?

One main reason is that the PFB approach is considerably more general, i.e. one can consider associated bundles for a wider variety of Lie groups; there is generally not a one-to-one correspondence between principal connections and connections on an associated bundle except in certain special cases which happen to include the standard model. Another likely reason, as described in Section~\ref{sec:PFB}, is that covariant derivatives on vector bundles are typically introduced as structures \emph{induced} by principal connections $\varpi$ on principal fibre bundles. $\varpi$ is (mistakenly)  taken to be more fundamental than any covariant derivative, $\D_i$ because it appears necessary to coordinate all the $\D_i$  across associated bundles. We have already refuted this reasoning in Section~\ref{sec:vb_A-B}. But there is another, related reason, that has come up in discussions of  \citep{Gomes_internal}. It is the assumption that in the absence of matter fields (i.e. sections of $E_i$), a covariant derivative $\D_i$ lacks physical meaning, whereas $\varpi$ has no such limitation, since it lives in a geometric structure that is independent of $E_i$. Hence, it is thought, any complete account of the theory—one that includes vacuum configurations—must appeal to the principal connection $\varpi$ (or its spacetime proxy $A_\mu^I$).

But this is no more compelling a reason than claiming that spacetime curvature in general relativity only acquires significance in the presence of particles. On the contrary, it is standard practice to describe spacetime geometry using an affine covariant derivative $\nabla$ on $TM$, irrespective of whether matter fields are present to ``feel'' that affine structure. Likewise, a vector bundle can be curved even in the absence of any particular sections. As \citet[p.~612]{Leeds1999} puts it:

\begin{quote}
Notice (here again there is an analogy with the affine connection in spacetime) the vector potential in any spatial region characterizes the region, and not what happens to be in the region; [...] So it is thought of as a field, one which, like many other fields, we describe in terms of what it acts upon. It plays a crucial role in the dynamics: in the A--B experiment, the phases of the wave at two different points on either of the paths enclosing the solenoid are related by parallel transport along that path; it is because of the path dependence of parallel transport that the two components of the wave acquire a phase difference.
\end{quote}

Now, in Section \ref{sec:vb_A-B} I quoted \cite{Jacobs_PFB} as defending the need for a \lq{}common cause\rq{} for the parallel transport over different vector bundles. In this regard, his main objection to the familiar account is that the principal connection inhabits a different space than the vector bundles: the principal connection is not a \lq{}matter field\rq{}; it is not a section of a vector bundle like other matter fields. Thus, he introduces into the philosophical literature the \emph{bundle of connections}. This is defined in much the same way as the associated bundle is defined from a vector space and a principal bundle

 Here I will proceed for left-invariant vector fields (i.e. those such that $L_g{}_*Z=Z$), but the analogous idea works for pseudo-tensorial forms.  Thus
\be(p, Z_p) \sim (g\cdot p, {L_g}_*(Z_p)), \quad \text{for all}\quad g\in G.\ee
Since locally (i.e. given some trivialization of the tangent  bundle) for $x=\pi(p)$ and $\xi\in \mathfrak{g}$, we can represent $p=(x, g):=g\cdot \sigma(x)$ and $Z_p=(X_{x}, \xi):=\xi+\sigma_*(X_x)$, where $X_x\in T_xM$, we have, locally, $(p, v_p)=(x,g, X_x, \xi)$. % and $(h\cdot p, h_*(v_p))=(x,hg, X_x, \Ad_h\xi)$. 
If we take the quotient,  elements of the new vector bundle will be locally of the form $(x, X_x, \xi)$, as was to be expected from a Lie-algebra valued 1-form (or vector field).  In other words,  if we know what parallel transport is at $p$, we know what it is at $g\cdot p$. By getting rid of this redundancy, we can find a global spacetime representation of the connection $\varpi$. This Atyiah-Lie connection is a section on the \textit{bundle of connections}, i.e. $\Upsilon\in \Gamma(T^*P/G)$, where $T^*P/G$ is a vector bundle over spacetime.\footnote{ See e.g. \cite[Sec. 3.2]{Ciambelli}; \cite[p.9]{LeonZajac}; \cite[p.60]{sardanashvily2009fibre}; \cite[Ch. 17.4]{Kolar_book} and \citep{Jacobs_PFB, Gomes_internal} for conceptual appraisals. The bundle of connections appeared almost simultaneously in  \cite{AtiyahLie} and  \cite{Kobayaschi_bundle}.  See also \cite[Ch. 17.4]{Kolar_book}. To avoid confusion, it is better to refer to a section of the bundle of connections, which is itself a generalization  of a connection to what are known as Lie algebroids (see \cite{mackenzie_2005}), as an Atiyah-Lie connection.} 

This construction suffices to have both matter fields and connections defined over the same space, as vector bundles, and it is an important step forward. However,  as argued in \citep{Gomes_internal}, having  a coordinate-independent representation of $\varpi$ on spacetime does not suffice to give \lq{}common cause\rq{} for the covariant derivatives $\D_i$ of the associated vector bundles: one must still stipulate that $\Upsilon$ induces the connection and curvature \emph{on the vector bundles of the other matter fields}. For $\Upsilon$ is still valued on the Lie algebra, and thus requires us to stipulate a representation so that it can act on matter fields. Of couse, there is no fundamental obstruction to this stipulation, just as there isn\rq{}t one in the textbook account:  in the course of  writing the dynamics these representations are posited without second thought. On the other hand, in the approach defended here, a fundamental vector bundle $E_i $ serves as that common cause, already at a kinematical or geometrical level, by fixing the relationship between all the induced covariant derivatives over its tensor products. 

\subsubsection{A local geometric effect}

Once one accepts the mistaken view that principal connections $\varpi$ are primary, she risks being misled about questions of locality—since  connections do not transform tensorially.

In more detail, gauge transformations are  ``vertical'' automorphisms acting fibrewise via group-valued functions $g: M \to G$ (see Equation \eqref{eq:vert_auto}). Under such transformations, $\varpi$ transforms \emph{inhomogeneously}, as per Equation \eqref{eq:varpi_inho}.  And since horizontal subspaces are defined by $\ker(\varpi)$, these inhomogeneous transformations generally fail to preserve horizontality. This is what makes a local interpretation of their effects challenging: at a point $p\in P$, any horizontal subspace can be taken to any other by a suitable gauge transformation; had they been tensorial, horizontal subspaces would be taken to horizontal subspaces.\footnote{ Though this holds for both Equation \eqref{eq:varpi_inho} and the version  for infinitesimal gauge transformations, $\xi:M\rightarrow \mathfrak{g}$, given in \eqref{eq:varpi_inf}, it may be clearest to see in the latter version: take $\mathbf{v}_h\in H_p\subset T_pP$ such that $\pi_*(\mathbf{v}_h)=\mathbf{v}\in T_xM$, then we get 
%\be \varpi(\mathbf{v}_h)=0\Rightarrow \widetilde{\varpi}(L_{g*}\mathbf{v}_h)=  \d g(\mathbf{v}) g^{-1}\ee  where $L_{g*}$ is the push-forward by the (left) action of the group element at (the fiber containing) $p$.
\be (\delta_{\xi(x)}\varpi)(\mathbf{v}_h)=\d\xi(\mathbf{v}).
\ee
 Or take Equation \eqref{ftnt:TE}: it is still true, and for the same reasons, that, at a point, $x\in M$, any non-vertical subspace of $TE$ can be horizontal. Of course, the connection represents  a \lq{}correction term\rq{} to the tangent of a vector field; it is what ensures the covariant derivative and parallel transport regain equivariance; see footnote \ref{ftnt:cov} }

Even sections of the bundle of connections, intended to address the redundancy of the principal fiber bundle, inherit the same property. Although $\Upsilon$  encodes $\varpi$ without reference to a local trivialisation,  it still transforms inhomogeneously. As \citet[p.~40]{Jacobs_PFB}  acknowledges:
 \begin{quote} Notice that the claim here is not that either a section [of the bundle of connections] or the connection are invariant under gauge transformations. This is not the case: a gauge transformation changes which direction on the bundle counts as `horizontal'. \end{quote}

Similarly, the gauge bosons $A_\mu^I$—often taken to be the theory’s dynamical variables—are not tensorial; they are not like the matter fields. Their transformation properties are also inhomogeneous, and thus involve derivatives.  Extracting what is left invariant under these transformations requires integrating over extended regions—that is, solving differential equations. This is what we do in the electromagnetic A-B effect, and it is the source of the concerns about non-locality.\footnote{And these concerns are borne out also more generally. One way to extract the invariant content from a given profile of $A_\mu^I$ over a given region (and not only along a curve) is to construct a `dressed observable' from gauge-fixed variables that depend solely on the gauge potential. And these are invariably non-local. By contrast, gauge-fixing procedures that rely on tensorial matter fields—such as the unitary gauge described in Section~\ref{sec:AB_EM}—yield observables that are entirely \emph{local} (see \citep{Rep_conv} for a detailed comparison).}

So here is the crux of this Section: gauge transformations do not preserve horizontality because the connection form \( \varpi \)—like the gauge potential \( A_\mu^I \), sections of the bundle of connections \( \Upsilon \), or the Christoffel symbols \( \Gamma_{ijk} \)—fails to transform tensorially, or more precisely, \emph{equivariantly}, under those transformations. This is an obstruction to an interpretation of these geometrical objects that is local, i.e. infinitesimal, in spacetime.  It thus makes sense  to take equivariance under general automorphisms to be the defining property of what I\rq{}ll call \emph{local geometric operators}; and \( \varpi , A_\mu^I,  \Upsilon\) don\rq{}t qualify.\footnote{One can, of course, still appeal to metaphysical doctrines—such as sophistication—to dismiss the physical significance of (the inhomogeneous) redundancy in $\varpi$ (and their descendants).  \cite{Jacobs_PFB} aptly defends internal sophistication for vertical automorphisms of a principal fiber bundle, but is unclear about whether he takes parallel transport to encode the dynamical content or $\Upsilon$ (or $\varpi$).}

%But why should equivariance be the criterion? The formalisation of geometric objects in this sense dates back at least to \citet{Nijenhuis_PhD}, and has been further developed in the recent literature (see e.g. \citep[Sec.~10.3]{Read_geom} for a conceptual overview). The key idea is that geometric objects are those whose coordinate representations are genuinely perspectival—different expressions of the same invariant structure. This perspectival coherence requires a strict formal condition: that the transition between representations composes correctly. In particular, transforming the representation of an object from chart \( A \) to chart \( C \) must yield the same result as transforming from \( A \) to \( B \), then from \( B \) to \( C \). 

%Objects that transform \emph{inhomogeneously}—such as \( \varpi \), \( A_\mu^I \), \( \Upsilon \), and \( \Gamma_{ijk} \)—fail to satisfy a generalised version of this condition. Their transformation behaviour under the automorphisms of the structure introduces history-dependence into what should be purely perspectival transitions. In this precise sense, they are not `geometric\rq{}.
% I believe internal sophistication in the case of principal fiber bundles is thus more complicated than it appears; I leave this topic for future work. } 

Alternatively, one might appeal to objects that transform equivariantly but are  extended,\footnote{See \citep{ButterfieldPoint} for a careful philosophical analysis of different senses of `extended'.} or to objects that are local and geometric but fail to capture the entire gauge-invariant content of the covariant derivative. These are, respectively, the cases of the holonomy (and, similarly, of dressed variables), and of the curvature. Yet both come with limitations: the former still sacrifices (some degree of) locality, while the latter omits global features. Neither fully resolves the conceptual puzzle posed by the Aharonov–Bohm effect, namely: how can local, i.e. only infinitesimally extended, interactions produce a globally detectable phase shift in the absence of local curvature?

By contrast, the covariant derivative \( \D \), which can also be defined from a connection on \( E \), is neither extended in the same sense as the holonomy or dressed variables, nor is it blind to global structure. It transforms according to the covariant rule given in \eqref{conex}, and thus qualifies as both local and geometric in the relevant sense.  For example, given $\mathbf{v}\in T_xM$, $\D_\mathbf{v}\kappa=0$ iff, the transformation under a linear automorphism of the vector bundle, $\widetilde{\D}_\mathbf{v}\widetilde{\kappa}=0$.\footnote{It is, as usual, the presence of the derivative acting on \( \kappa \), together with the transformation properties of \( \hat V \), that ensures the covariant transformation behaviour of \( \D \) as defined by the connection; see Equation~\eqref{ftnt:TE}.\label{ftnt:cov}}

The view defended here is thus that both (a) and (b) are best answered via covariant derivatives on fundamental vector bundles. This dissolves many of the apparent puzzles surrounding locality and invariance in the generalised, classical Aharonov–Bohm effect.

% Sections of the bundle of connections, $\Upsilon$, are not geometric in the sense described above either, for they still transform inhomogeneously. Of course we can appeal to derived mathematical objects, such as parallel transport, dressed fields,  or the curvature.%\footnote{Though nothing in my argument runs on this, note that, given  $\varpi$ on $P$, and $E$ seen as an associated vector bundle to $P$, $\varpi$ uniquely induces a covariant derivative $\D$ on $E$. Moreover, on a general vector bundle $E$, a covariant derivative $\D$ uniquely gives a notion of parallel transport and vice-versa. But the infinitesimal notion of parallel transport  on a vector bundle gives rise to a covariant derivative again, so parallel transport on the vector bundle doesn\rq{}t \emph{uniquely} induce a principal connection $\varpi$: ultimately, there is no canonical isomorphism between $\mathsf{Aut}(E)\subset\mathsf{End}(E)$ and the Lie algebra $\mathfrak{g}$. Given a representation of $\mathfrak{g}$ on the typical fiber $V$, there is unique isomorphism between $\mathsf{Aut}(V)$  and $\mathfrak{g}$, but there is no unique isomorphism between $\mathsf{Aut}(V)$ and $\mathsf{Aut}(E)$; this would require a section of $E$ or a choice of flat derivative,  as described in Equations \eqref{eq:omega} and \eqref{eq:D}. Alternatively, $\D$ uniquely defines a connection on $TE$, as the complement of the canonical vertical space $VE$; see Equation \ref{ftnt:TE}.} 

\section{Summary}\label{sec:conclusions} 

In light of the classical A-B effect, do we need to revise gauge theory? This is a popular conclusion.  

Let me again be clear that the observable effect as a whole---namely the total phase shift---causes no trouble for any picture of gauge theory that awards physical significance to those and only those quantities that are gauge-invariant; the total phase is such a quantity. None of the revisions aims to challenge the physical significance of the total shift, nor that it is gauge-invariant. 
Moreover, all such revisions eschew an action at a distance from the magnetic fields right from the bat. 

The real motivation, the main aim of revision, is to either dismiss or give a plausible account of the infinitesimal \emph{accrual} of the total phase shift. Whether an electron gains a phase shift of $\theta$ or $\theta\rq{}$ in a proper segment of $\gamma_1$ or $\gamma_2$ depends not on the local magnetic field, but on whether we use $\mathbf{A}$ or its gauge-related $\mathbf{A}\rq{}$ to describe the effect. How can we understand this indeterminacy through standard gauge theory, which is usually taken to  afford physical significance only to gauge-invariant quantities? 

Some have used the puzzles posed by the effect to defend an alternative set of variables for the theory  (see  \cite{Healey_book}  and references therein). The \lq{}holonomy-interpretation\rq{} of gauge theory, defended most vigorously by \citet{Healey_book}, falls under this kind of revision. Here, there just is no local accrual of the total phase shift; it cannot even be formulated using the holonomy variables. But this interpretation has many flaws: it is non-local (both synchronically or diachronically, in the terminology of \cite{Belot1998}), and one cannot write the equations of motion using the holonomy variables (cf. \cite[Sec. 3.5.2]{Gomes_PhD} for a summary of these and other flaws). 

Others have argued that a particular choice of gauge is ontologically preferred, the \lq{}preferred gauge\rq{} (cf. \citep{Maudlin_response, Maudlin_ontology, Mattingly_gauge, Mulder_AB, Gao_AB}). This kind of revision seeks to maintain locality and give a plausible account of the local accrual of the phase, in one form or another, arguing that there is a fact of the matter about which member of any family of gauge-related distributions truly \lq{}refers\rq{}, with others being allowed only by a kind of empirical underdetermination. Thus \cite{Wallace_deflating, Wallace_unitary} defends unitary gauge; \cite{Maudlin_ontology} defends Coulomb gauge; \cite{Mattingly_gauge, Mulder_AB, Gao_AB} defend Lorenz gauge, etc.\footnote{I count the treatments of \cite{Leeds1999} and \cite{Jacobs_PFB} to be the closest to my proposal here, though details differ greatly.} 

Here, I align with the first kind of revisionist in rejecting the significance of the local accrual of the A-B phase and with the second in maintaining locality.\footnote{With the exception of \cite{Maudlin_ontology}.}   But I depart from the first by ensuring my variables \emph{can} express the dynamics of Yang-Mills theory and the local accrual of phase---though this accrual vanishes in the A-B effect. And I depart from the second by foregoing any gauge choice altogether.

To walk that narrow path between the two types of revision, I employed a formalism of gauge theory that trades the gauge potential (or the principal connection form in a principal bundle) acting on associated vector bundles for a covariant derivative on a fundamental vector bundle, with other associated vector bundles built as tensor products.  I called this the vector bundle point of view, and I believe it clarifies certain questions of ontology, such as that of the `coordination problem\rq{}, as discussed by e.g. \cite{Jacobs_PFB}. 

 The ontology suggested by this alternative formalism is local, and very similar to that defended by  \cite{Leeds1999}. Indeed, it is also similar to the ontology proposed by \cite{Wallace_deflating}\rq{}s intepretation of unitary gauge. Namely, particles have properties represented by vectors in spaces that are attached to spacetime points, and the gauge potential stands for an infinitesimal comparison of values of these vectors at neighboring points, a comparison that is geometrically captured by the covariant derivative $\D$.   That is, $\D$ determines how elements of $E_x$ are related to elements of $E_{x+\delta x}$; though $\D$ is a legitimate object even in the absence of matter fields (sections of $E$), just like the spacetime covariant derivative $\nabla$ is legitimate without there being matter fields to be acted upon by it. 
 
% Moreover, as \citet[Sec. 3.3]{Jacobs_PFB} points out,  Wallace\rq{}s interpretation falters once we include more than one matter field: which of these bear the relations that define the connection? More than one field charged under the same force, would, according to Jacobs, lead to an underdetermination in the ontological picture as proposed by Wallace. But unlike Wallace\rq{}s, my picture has a fundamental vector bundle whose covariant derivative is defined independently of the consideration of any concrete vector field thereupon. %I don\rq{}t take this picture to point to the inadequacy of a vacuum sector of the theory, i.e., one that doesn\rq{}t include sections of $E$ or its tensor products. 

 This is not the only analogy between gauge theory in the vector bundle formalism and general relativity; they are structurally identical in many respects. In the vector bundle point of view, it is clear that we should understand the A-B effect precisely as we understand the disagreement between a spacetime vector at $x$ and its parallel transport around a loop.  That means we cannot infer the total disagreement from the local accrual, or intrinsic local rotation, for that vanishes under parallel transport. Upon reunion, the two parallel transported vectors can simply disagree, although neither has locally rotated: this is what happens in the A-B effect.  This conclusion can also be a sticking point for the student of differential geometry, but it would be a mistake to think it points to a fundamental explanatory gap.

%In both the general relativistic holonomy and the A-B effect, there seems to be something holistic at play: although the total disagreement is measurable, there simply is no fact of the matter as to how it comes about as the result of small, locally accrued differences. These are just facts about parallel transport that are evinced only globally, or rather, only when the paths re-converge.   

Of course, instead of intrinsic rotation, we can talk about rotation relative to a coordinate chart, in which case rotation would be encoded by the Christoffel symbols of the covariant derivative as expressed in that chart. %But it goes without saying that such an extrinsic notion of rotation is largely arbitrary and purely representational.
\emph{Mutatis mutandis} for the gauge case:  in our intrinsic description in terms of a vector bundle and its covariant derivative, $\D$, the dependence on a choice of gauge only appears if we use $\mathbf{A}$ or $\omega$, representatives of $\D$ relative to a flat connection or a local trivialisation of $E$, respectively.  And using $\mathbf{A}$ without keeping track of its origin puts us back under the by now familiar misconceptions surrounding the A-B effect.  
 
The last criticism I fended off in this paper regarded the active automorphisms of the structures of the theory, as opposed to its passive coordinate changes.\footnote{This is a criticism that \citet{Leeds1999} leaves unanswered, when he says that his picture (ibid. 613) \lq\lq{}traffics heavily in non-measurable properties and quantities”.} My defense was that $\D$ is a local geometric operator, i.e. one that transforms equivariantly under the linear automorphisms of the vector bundle. If a vector field has vanishing covariant derivative at a point, it will transform to one with the same property under an automorphism.  Thus, I again see no difference between the redundancy due to the automorphisms of a vector bundle and the one familiar from differential geometry, where  the diffeomorphisms of the background smooth spacetime manifold act non-trivially but equivariantly on tensors. The go-to difference between the two cases---the \lq{}soldering\rq{} of spacetime tensors---is not sufficient to eliminate the \lq{}active\rq{} redundancy in the holonomy of spacetime vectors either.  And yet this lack of invariance doesn\rq{}t cast doubt on the idea and significance of non-trivial holonomies for spacetime vectors; the same goes here.% Let us not stack up unrelated puzzle  the hole argument in general relativity (cf. e.g. \citep{Pooley_Read, GomesButterfield_hole1} for recent reviews)  

Lastly, let me be clear:  I \emph{am} saying that the major puzzle at the center of most discussions of the A-B effect dissolves into familiar concepts from differential geometry, and that this is most clearly seen in the vector bundle formalism.   But I am \emph{not} saying that the A-B effect has nothing to teach us beyond the familiar lore of differential geometry. No, the effect still packs a striking lesson: that, even jointly, the local curvature and topology may fail to determine the physical content of the local covariant derivative. This idea is encountered in differential geometry and we should adopt it equally in gauge theory.\footnote{Although it is often said that general relativity is a theory about the curvature of spacetime, by which one means a theory of the Riemann curvature tensor, and this description is mostly accompanied by a proviso regarding different topologies, more careful treatment acknowledge that even that may be insufficient to determine the full physical content of the theory: this is given by the metric, which recovers the Levi-Civita covariant derivative.} In Abelian gauge theories, this underdetermination occurs only in non-simply connected regions; in non-Abelian theories, it occurs without any topological constraints.  Thus, rather than dissolving into the familiar lore of differential geometry, the A-B effect may offer it new insights. 
  
 % To close off, let me mention a disanalogy between the gravitational and the gauge A-B effects that remains. Namely,  that due to shielding and universality. We can, in the laboratory, easily shield magnetic sources from the paths of the electron. Shielding gravitational sources, however, is not so easy.\footnote{ See \cite{ChruscielBeig_shield, CarlottoSchoen2016} for some interesting shielding results in perturbative and non-perturbative gravity, respectively. } For that, we may need more esoteric gravitational objects, such as cosmic strings. But  I don\rq{}t take this feature as germane to the comparison between the gravitational or spacetime and the gauge A-B effect. Our analysis was mostly kinematical, and this difference is mostly about dynamics. 
  
%Summing up, apart from differences that are due to dynamical features of gravity (such as the absence  of particles with negative mass that would allow shielding), there seems to be no conceptual distinction between the Aharonov-Bohm effect and parallel transport  in an arbitrarily curved spacetime. %Indeed, we can even get a non-trivial holonomy of spacetime vectors in regions with zero curvature, by employing conical singularities. In both cases, one cannot, by surveying a neighborhood of the particles' trajectories, infer whether they will experience a relative shift when they are once again reunited. In both cases, no mystery remains. 
 
 \subsection*{Acknowledgements}
 
 I would like to thank the British Academy for financial support and Caspar Jacobs for feedback. 
    \begin{appendix}

   \section*{APPENDIX}
   \section{ Vector, principal,  and associated fiber bundles}\label{app:PFB}
      
   \begin{defi}[Principal Fiber Bundle] $(P, M,G)$ consists of a smooth manifold $P$ that admits a smooth free action of a  (path-connected, semi-simple) Lie group, $G$: i.e.  there is a map $G\times P\rightarrow P$ with $(g,p)\mapsto g\cdot p$ for some left action $\cdot$ and such that for each $p\in{P}$, the isotropy group is the identity (i.e. $G_p:=\{g\in{G} ~|~ g\cdot p=p\}=\{e\}$). $P$ has a canonical, differentiable, surjective map, called a projection, under the equivalence relation $p\sim g\cdot p$, such that $\pi:P\rightarrow P/G\simeq M$, where here $\simeq$ stands for a diffeomorphism. \label{def:PFB}\end{defi}
It follows from the definition that $\pi^{-1}(x)=\{G\cdot p\}$ for $\pi(p)=x$.  And so there is a diffeomorphism between $G$ and $\pi^{-1}(x)$, fixed by a choice of $p\in \pi^{-1}(x)$. It also follows (more subtly) from the definition, that local sections of $P$ exist.  Similarly to a section of $E$, a local section of $P$ over $U\subset M$ is a map,  $\sigma: U\rightarrow P$ such that $\pi\circ\sigma=\mathrm{Id}_U$. Unlike sections of vector bundles, sections of principal bundles are generally only local. 
 
 The automorphism group of a principal bundle $P$ consists of fibre-preserving diffeomorphisms: 
   \begin{equation} \tau: P \to P \quad \text{such that} \quad \tau(g \cdot p) = g \cdot \tau(p). \label{eq:vert_auto}\end{equation} 
   Gauge transformations are a distinguished subclass: those for which $\pi \circ \tau = \pi$, i.e., ``vertical'' automorphisms acting fibrewise via group-valued functions $g: M \to G$. 
   
  Given an element $\xi$ of the Lie-algebra $\mathfrak{g}$, and the action of $G$ on $P$, we use the exponential to find an action of $\mathfrak{g}$ on $P$. This defines an embedding of the Lie algebra into the tangent space at each point, given by the \emph{hash} operator: $\#_p: \mathfrak{g}\rightarrow T_pP$. The image of this embedding we call \emph{the vertical space} $V_p$ at a point $p\in P$:  it is tangent to the orbits of the group, and is linearly spanned by vectors of the form 
\be\label{eq:fund_vec} \text{for}\quad \xi\in \mathfrak{g}:\quad {\xi^\#}(p):=\frac{d}{dt}{}_{|t=0}(\exp(t\xi)\cdot p)\in V_p\subset T_pP.\ee
Vector fields of the form $\xi^\#$ for $\xi\in \mathfrak{g}$ are called \emph{fundamental vector fields}.\footnote{It is important to note that there are vector fields that are vertical and yet are not fundamental, since they may depend on $x\in M$ (or on the orbit).  \label{ftnt:vertical}} 

 The vertical spaces are defined canonically from the group action, as in \eqref{eq:fund_vec}.  But we can define an \lq{}orthogonal\rq{} projection operator, $\hat V$ such that: 
   \be\label{eq:ortho_H} \hat V|_V=\mathsf{Id}|_V,  \quad \hat V\circ \hat V=\hat V,\ee 
   and defining $H\subset TP$ as $ H:=\mathsf{ker}(\hat V)$. It follows that $\hat H=\mathsf{Id}-\hat V$ and so $\hat V\circ \hat H=\hat H\circ \hat V=0$.
     Moreover, since $\pi_*\circ \hat V=0$ it follows that:
   \be \pi_*\circ \hat H=\pi_*.
   \ee
   
The connection-form should be visualized essentially as the  projection onto the vertical spaces: given some infinitesimal direction, or change of frames, the vertical projection picks out the part of that change that was due solely to a translation across the group orbit.  %This ensures that, for any $p\in \pi^{-1}(x)$, $\pi_*H_{p}\simeq T_xM$. 
The only difference between $\hat V$ and $\varpi$  is that the latter is $\mathfrak{g}$-valued, Thus we get it via the isomorphism between  $V_p$ and $\mathfrak{g}$ ($\varpi$'s  inverse is $\#: \mathfrak{g}\mapsto V\subset TP$).  %by the first condition of \eqref{eq:omega_defs}  we can also define a horizontal space as the kernel $\mathsf{Ker}(\varpi_p)=:H_p$. That is,  since $\varpi_p{}|_{V_p}$ is a linear isomorphism, we get that $\mathsf{Ker}(\varpi_p)$ and  $V_p$ are transversal and $\mathsf{Ker}(\varpi_p)\oplus V_p=T_pP$.\footnote{More directly: \begin{align*}\text{dim}(T_pP)&= \text{dim}(\mathsf{Ker}(\varpi_p))+\text{dim}(\mathsf{Im}(\varpi_p))\\\, &= \text{dim}(T_xM)+\text{dim}(\mathfrak{g}),\end{align*} and $\text{dim}(\mathsf{Im}(\varpi_p))=\text{dim}(\mathfrak{g})$, we obtain $\text{dim}(\mathsf{Ker}(\varpi_p))=\text{dim}(T_xM)$, with $\pi_*$ an isomorphism between the two.} 
%Thus the vectors spanning $\mathsf{Ker}(\varpi_p)=\mathsf{Ker}(\hat V_p)$ can be defined as the  {horizontal} vectors in the bundle, and each represents a unique `horizontal lift' at $p$ of a direction at $T_{x}M$.  The second condition of \eqref{eq:omega_defs} guarantees that the notion of horizontality covaries with the choice of representative of the fiber (e.g. the choice of frame in the frame bundle example above), that is: a vector $v\in T_pP$ is horizontal iff ${L_g}_* v\in T_{g\cdot p}P$ is horizontal. %It is this kind of covariance that allows us to see the action of the symmetry group as  representing a kind of `redundancy'. 
 We can define it directly as:
\begin{defi}[An principal connection-form]  $\varpi$ is defined as a Lie-algebra valued one form on $P$, satisfying the following properties:
\be\label{eq:omega_defs}
\varpi(\xi^\#)=\xi
\qquad\text{and}\qquad
{L_g}^*\varpi=\Ad_g\varpi,
\ee
%\begin{eqnarray}
%\varpi(v_\xi)&=&\xi\notag\\
%{L_g}^*\varpi&=&g^{-1}\varpi g\notag
%\end{eqnarray}
  where the adjoint representation of $G$ on $\mathfrak{g}$ is defined as $\Ad_g\xi=g\xi g^{-1}$, for $\xi\in \mathfrak{g}$;  ${L_g}^*$ is the pull-back of $TP$ induced by the diffeomorphism  $g:P\rightarrow P$.\end{defi}

  Under the vertical automorphisms, \eqref{eq:vert_auto}, $\varpi$ transforms \emph{inhomogeneously}: 
\begin{equation} \widetilde{\varpi} = (dg)g^{-1} + g\varpi g^{-1},\label{eq:varpi_inho} \end{equation} 
Infinitesimally, i.e. for Lie-algebra-valued spacetime-dependent transformations $\xi:M\rightarrow \mathfrak{g}$, 
\be \label{eq:varpi_inf}
\delta_{\xi(x)}\varpi=[\varpi, \xi(x)]+\d \xi(x),
\ee
where, reinstating the $\pi(p)$ in place of $x$, we read the action of the second term on $Z\in \Gamma(TP)$ as 
$ \d\xi(\pi(p))(Z)=\pi_*(Z)[\xi(\pi(p))]$, which, in a local trivialisation takes the derivative of the spacetime function and leaves the Lie-algebra values intact 
(see \citep[Ch.~II]{kobayashivol1} or \citep[Eq.~3.2.11]{Gomes_elements}).

   Now, in possession of an principal connection, we can induce a notion of covariant derivative on \emph{associated vector bundles}:
 \begin{defi}[Associated Vector Bundle] A   vector bundle over $M$ with typical fiber $V$, is associated to $P$ with structure group $G$, is defined as: 
\be E=P\times_\rho V:=P\times V/\sim\quad \text{where}\quad (p,v)\sim  (gp, \rho(g^{-1})v),\label{eq:AVB}
\ee
where $\rho:G\rightarrow GL(V)$ is a representation of $G$ on $V$.\label{def:AFB}\end{defi}

 Given any vector bundle $(E, M, V)$, the bundle of frames for $E$, called $L(E)$, is itself a principal fiber bundle $(L(E), M,GL(V))$: here elements of $\pi^{-1}(x)$ are linear frames of $E_x$, and   $G\simeq GL(F)$ acts via $\rho$ on the typical fibers. By construction, $E\simeq L(E)\times_\rho V$.    If $V$ has more than just the structure of a linear vector space, e.g. if it is endowed with an inner product, then we have \emph{bundle of admissible frames}, e.g. orthonormal frames. This is also a principal fiber bundle, $(L\rq{}(E), M, G)$, 
whose structure group is a proper subgroup of the general linear group, $G\subset GL(V)$, taken to be the group that preserves the structure of $V$.

One can get a  covariant derivative on an associated vector bundle $E$ from  $\varpi$  as follows: 
let $\gamma:I\rightarrow M$ be a curve tangent to $\mathbf{v}\in T_xM$, and consider its horizontal lift, $\gamma_h$. Suppose $\kappa(x)=[p, v]$. Then 
\be
\nabla_\mathbf{v}\kappa=\frac{d}{dt}[\gamma_h, v].
\ee

Conversely, we can define a horizontal subspace from the covariant derivatives as follows. For $p={e_1,...e_n}\in L(E)$, and for all curves $\gamma\in M$ such that $\mathbf{v}=\dot\gamma(0)\in T_xM$, with $\pi(p)=x$, let $\{e_1(t),..., e_n(t)\}$ be curves in $E$ such that $\nabla_\mathbf{v}(e_i(t))=0$. Doing this for each $v$ defines a horizontal subspace.

\subsubsection*{Covariant derivative as an operator on $TE$}
Indeed, given any vector bundle, we have a similar definition of covariant derivative that bypasses the principal bundle formalism, as in \eqref{eq:affine_E}. In other words, following the idea that a connection should relate elements of neighboring fibers, we label as \lq{}vertical\rq{}  the tangent space to $\pi_E^{-1}(x)$, i.e. the tangent space to $E_p$, seen as a subspace of $TE$ (generated by curves in $E_p$), which is also canonical. So here too, a connection is given by a projection operator as in \eqref{eq:ortho_H}: $\hat V: TE\rightarrow TE$, onto the vertical subspace, $V\subset TE$. We can then define a covariant derivative straightforwardly as follows. Take $\kappa\in \Gamma(E)$ and a curve  $\gamma:I\rightarrow M$ tangent to $\mathbf{v}\in T_xM$. Then $\kappa|_\gamma$ is a curve in $E$, with tangent $\dot\kappa$. We  then define
\be \nabla_\mathbf{v} \kappa:=\hat V(\dot\kappa).\label{ftnt:TE}\ee
It is the extra derivative on $\kappa$, jointly with the transformation property of $\hat V$, that ensures the transformation property of $\nabla$ is covariant, i.e. given by \eqref{conex}. 
\end{appendix}
\bibliographystyle{apacite} 
\bibliography{references3}

\end{document}